\documentclass[
aps,nofootinbib,
showpacs,showkeys,preprint
tightenlines,preprintnumbers,] {revtex4}

\usepackage{epsf,epsfig,subfigure,graphicx,amsmath,amssymb 
}
\usepackage{color}
\newcommand{\dis}[1]{\begin{equation}\begin{split}#1\end{split}\end{equation}}
\newcommand{\ie}{{\it i.e.~}}

\newcommand{\Mg}{{M_{\rm GUT}}}

\def\sw0{{$\sin^2\theta_W^0$}}

\newcommand{\Z}{{\bf Z}}

\def\E6{{\rm E_6}}

\def\EE8{{\rm E_8\times E_8'}}
\def\GG{SU$(5)_{\rm GG}$}
\def\flip{SU$(5)_{\rm flip}$}
\def\anti{anti-SU$(5)$}
\def\antiFD{SU$(5)_{\rm anti2}$}
\def\antiSD{SU$(7)_{\rm anti2}$}

\def\antinD{SU$(N)_{\rm anti2}$}
\def\antinT{SU$(N)_{\rm anti3}$}
\def\antinQ{SU$(N)_{\rm anti4}$}

\def\four{{\bf 4}}

\def\one{\bf 1}

\def\five{\bf 5}
\def\ten{\bf 10}
\def\tenb{\overline{\bf 10}}

\def\fiveb{\overline{\bf 5}}

\begin{document}

\draft

\title{\Large\bf Towards  unity of families: anti-SU(7) from $\Z_{12-I}$ orbifold compactification}

\author{ Jihn E.  Kim}
\address
{Department of Physics, Kyung Hee University, 26 Gyungheedaero, Dongdaemun-Gu, Seoul 02447, Republic of Korea, and\\
Center for Axion and Precision Physics Research (IBS),
  291 Daehakro, Yuseong-Gu, Daejeon 305-701, Republic of Korea 
}

\begin{abstract} 
The problem of families, ``Why are there three families of fermions?'', is a long awaited question to be answered within a reasonable framework. We propose anti-SU($N$) groups for the unification of families in grand unification (GUT) groups, where the separation of color and weak gauge groups in the GUT  is achieved by antisymmetric tensor Brout-Englert-Higgs boson instead of an adjoint representation. Theories of anti-SU($N$)'s are proposed for the unification of families. The minimal model is found as \antiSD \, GUT with the fermion representation $[\,3\,]+2\,[\,2\,]+8\,[\,\bar{1}\,]$. We present an example in  a $\Z_{12-I}$ orbifold compactification, where  the missing partner mechanism is also realized. 
 
\end{abstract}
\maketitle


\section{Introduction}\label{sec:Introduction}
   
The family problem, repeating fifteen chiral fields three times in the standard model (SM), has been known for almost four decades without an accepted theory so far. One family consists of fifteen chiral fields. In the Georgi-Glashow(GG) grand unification(GUT) model, the  fifteen chiral fields in the first family are grouped into ${\ten}\,(u^c,u,d,e^+)$ and $\fiveb\,(d^c,\nu_e,e)$ of \GG\,\cite{GG74}. The family problem posited in GUTs is how $\ten+\fiveb$ repeats exactly three times in an extended GUT. The gauge coupling unification \cite{GQW74} is the underlying principle for a GUT. A gauge model based on a simple group is a GUT.  A gauge model based on a semi-simple group rendering one coupling constant (by some  discrete symmetry) is a GUT also \cite{PS73}.

A chiral model with fifteen fields in a GUT is achieved by assigning fifteen in some complex representation of a GUT gauge group. Toward a solution of the family problem, Georgi formulated a principle on the unification of GUT-families with gauge groups containing SU(5) as a subgroup  \cite{Georgi79}. Absence of gauge anomaly is required. Then, any gauge group except SU($N$) can be used for a GUT if they allow complex representations. In this try of anomaly-free groups, SO($2N$) with $N=2n+1$ allows complex representaions which are spinors of SO($2N$). The simplest case is SO(10) where the fifteen chiral fields plus a singlet neutrino are assigned in the spinor {\bf 16}  \cite{SO10}. It is one family model. One may try SO(12) which however is not considered to be complex because the spinor {\bf 32} branches to ${\bf 16}$ and $\overline{\bf 16}$ of SO(10). So, the next step is a two family model where the  spinor {\bf 64} of  SO(14) is used \cite{Kim80,Kim81} where however non-standard charges must appear.\footnote{There can exist a  missing partner mechanism in this model, however, with certain assumptions.  This was emphasized in \cite{Dimopoulos81}. Later, it  was worked out in SO(10) \cite{Missing}, which is basically \anti=\flip.} If we exclude the possibility beyond the SM charges,  {\bf 64} of  SO(14) is vectorlike and no chiral family is obtained. A scheme toward  unification of GUT-families (UGUTF) came to a dead end within SO($2N$) groups.

To open a gate from the dead end alley toward the meadow, Georgi proposed UGUTFs in SU($N$) groups \cite{Georgi79}, where the condition on the anomaly-freedom plays a central role. There have been some attempts along this line \cite{Frampton79,FramptonPRL79}.
After the string revolution with heterotic strings \cite{Gross85}, the low energy gauge groups and spectra are computed through the compactification process \cite{Candelas85,Dixon85,Ibanez87}, and UGUTFs did not attain much interest because three SM families could have been obtained through the compactification process \cite{IKNQ87}. In this string scenario, the compactification schemes (such as orbifolds) basically choose what is the number of families. The drawback is that there are too many parameters to predict the fermion mass spectra.

In this paper, we attempt to realize Georgi's UGUTFs in string compactification. If possible, we will try to introduce some structure among  three families so that the difference of the third family from the the first two is understood.
Among heterotic strings, usually the $\EE8$ hetrotic string has been used because it contains spinors in the adjoint representation {\bf 248} of E$_8$. Then, the $\EE8$ heterotic string has been favored subconciously because of the embedding chain of the SO(10) spinor {\bf 16} in   SO(10)$\to$E$_6 \to$E$_7\to$E$_8$.  But, the SO(32) hetrotic string is also useful for phenomenology as we will briefly argue in this paper. The most severe obstacle in obtaining  a realistic SU(5) GUT from heterotic string has been the difficulty, at the level 1 construction, of obtaining the adjoint representation {\bf 24} of SU(5) which is needed to break SU(5) down to the SM gauge group.

The first example without an adjoint representation for the Brout-Englert-Higgs(BEH) boson can be traced back to anti-SU(7) where antisymmetric tensor fields for BEH bosons, denoted as two and three index anti-symmetric tensor fields, $\Phi^{[\alpha\beta]},\,\Phi_{[\alpha\beta]},\,\Phi^{[\alpha\beta\gamma]}$, and $\Phi_{[\alpha\beta\gamma]}$, are used to reduce the rank of the SU(7)$\times$U(1) GUT gauge group and separate the color and the weak parts \cite{Kim80,Kim81}.\footnote{The acronym anti- was used in Ref. \cite{DKN84} as  anti-SU(5).}
The sixteen chiral fields of the first family are grouped into ${\ten}_1\,(d^c,u,d,N_{1}^0)$ and $\fiveb_{-3}\,(d^c,\nu_e,e)$ and $e^+_5$ in \flip. The structure is included in \antiSD. 
If it is applied to SO($2N$) gauge groups, we can call it anti-SU($N$) where anti means that separation of color SU(3)$_c$ from weak SU(2)$_W$ is by the anti-symmetric tensor fields instead of the adjoint representation. For $N=5$, it is now known as the flipped SU(5) \cite{Barr82}. It is obvious that the essential feature is included in the {\it word}  `anti-SU($N$)'. Anti-SU(5) GUTs were obtained in string compactification \cite{Ellis89,KimKyae07}. But, SU(5) GUTs are not UGUTFs.

In addition to the restriction on the number of families for $n_f=3$, the R-parity is used for proton longevity and weakly interacting massive particle possibility for cold dark matter candidate. The R-parity can be a discrete subgroup of a U(1) gauge group which has been discussed in Ref. \cite{KimPRL13,KimPLB13}. Another issue in supersymmetric GUTs is the problem of doublet-triplet splitting in the BEH multiplets containing $H_u$ and $H_d$. We will realize the doublet-triplet splitting mechanism in \antiSD~anticipated in Ref. \cite{Dimopoulos81}.
   
In Sec. \ref{sec:UGUTF}, we recapitulate the old UGUTF scenario, and continue in Sec. \ref{section:Orb} to present a rationale that anti-SU($N$)s are theories for UGUTFs.
In Sec. \ref{sec:SU7Model}, we summarize an SU(7) realization of UGUTF in the $\Z_{12-I}$ orbifold compactification. Here, we present some details on how the massless spectra are obtained in $\Z_{12-I}$ so that line by line can be followed up in other orbifold constructions. We also point out how the missing-partner mechanism is realized in \antiSD.
Sec. \ref{sec:Conclusion} is a conclusion.  In Appendix, we present the spectra not included in Sec. \ref{sec:SU7Model}. 
 
\section{Families unified in grand unification}\label{sec:UGUTF}
 
Let the fundamental representations (or anti-symmetric representations)  of SU($N$) are bounded by square brackets. Representations [1] and [2] have  the following matrix forms,
\dis{
[1]\equiv \Phi^{[A]}=\begin{pmatrix}
\alpha_1\\[0.2em] \alpha_2\\[0.2em]  \alpha_3\\[0.2em]  \alpha_4\\[0.2em]  \alpha_5\\[0.2em]  f_6\\ \cdot \\[0.2em]  \cdot\\[0.2em]    \cdot\\[0.3em]  f_N\\[0.2em]
\end{pmatrix},\hskip 0.5cm
[2]\equiv \Phi^{[AB]}=\begin{pmatrix}
0, &\alpha_{12}, &\cdots,&\alpha_{15} & {\Big|}&\epsilon_{16},&\cdots,&\epsilon_{1N}\\[0.2em]
-\alpha_{12},& 0,&\cdots,&\alpha_{25}&{\Big|}&\epsilon_{26},&\cdots,&\epsilon_{2N}\\ 
\cdot&  \cdot&  \cdot&  \cdot&{\Big|}&  \cdot& \cdot& \cdot \\ 
\cdot&  \cdot&  \cdot&  {\color{red}\alpha_{45}}&{\Big|}&  \cdot& \cdot& \cdot \\ 
-\alpha_{15}, &-\alpha_{25},&  \cdots,&0&{\Big|}&\epsilon_{56},&\cdots,&\epsilon_{5N}\\[0.2em]
\hline
-\epsilon_{16},& -\epsilon_{26}\,&\cdots,&-\epsilon_{56}&{\Big|}&0,&\cdots,&\beta_{6N}\\ 
\cdot&  \cdot&  \cdot&  \cdot&{\Big|}&  \cdot& \cdot& \cdot \\ 
-\epsilon_{1N},& -\epsilon_{2N}\,&\cdots,&-\epsilon_{5N} &{\Big|}&-\beta_{6N},&\cdots,&0
 \end{pmatrix} \label{eq:FundSUN}
 } 
 where [1] means the dimension {\bf N}, $ [2]$ means the dimension $\left({\bf \frac{N(N-1)}{2!}} \right)$ with two antisymmetric indices,  $ [3]$ means the dimension $\left({\bf \frac{N(N-1)(N-2)}{3!}} \right)$ with three antisymmetric indices,  etc. We do not consider the symmetric indices such as $\{ 2 \}$ $\left(={\bf \frac{N(N+1)}{2!}} \right)$ because they will contain color sextets.  
[1] contains one $\five$, and [2] contains one $\ten$ of SU(5). The number of the SU(5)$_{\rm GG}$ families, \ie that of $\ten$ plus $\fiveb$, is counted by the number of $\ten$ minus the number of $\tenb$. 
The anomaly-freedom condition chooses the matching number of $\fiveb$'s. The numbers $n_1$ and $n_2$ for the vectorlike pairs $n_1(\five\oplus\fiveb)+ n_2(\ten\oplus\tenb)$ are not constrained by the anomaly freedom. Thus, we count the number of families  just by the net number of two index fermion representations in the SU(5)$_{\rm GG}$ subgroup.
  For several SU($N$)'s, we have the following family number $n_f$ by counting the number of $\ten$'s,
\dis{
&{\rm SU(5)}: [2]\to n_f=1\\
&{\rm SU(6)}: [3]\to n_f=0, ~[2]\to n_f=1\\
&{\rm SU(7)}: [3]\to n_f=1, ~[2]\to n_f=1\\
&{\rm SU(8)}: [4]\to n_f=0, ~[3]\to n_f=2, ~[2]\to n_f=1\\
&{\rm SU(9)}: [4]\to n_f=5, ~[3]\to n_f=3, ~[2]\to n_f=1\\
&{\rm SU(11)}: [5]\to n_f=-5, ~ [4]\to n_f=9, ~[3]\to n_f=5, ~[2]\to n_f=1 \label{eq:NumberGGf}
}
from which we define $[\overline{m}]=[N-m]$.
Negative $n_f$ gives the chiral family number $|n_f|$, which is also allowed.
The anomaly units in SU($N$) are
\dis{
 &~{\cal A}([m])= \frac{(N-3)!(N-2m)}{(N-m-1)!(m-1)!},\\ 
 &~{\cal A}({[1]})=1,~{\cal A}\left({ [2]}  \right)=N-4,~ {\cal A}\left([3] \right) = \frac{(N-3)(N-6)}{2},\,{\rm etc.} \label{eq:anomaly}
 } 
where $m\ge 1$. For $[\bar{m}]$, we use ${\cal A}([\bar{m}])=- {\cal A}([m])$.
From now on, we frequently use the indices to represent the antisymmetric $[m]$ in terms of $\Psi^{[\alpha_1,\cdots,\alpha_m]}$ where $\alpha_1,\alpha_2,\cdots,$  and $\alpha_m$ are completely antisymmetrized.

Theory of families in GUTs does not allow repetition of the representation and gauge anomalies \cite{Georgi79}.  I. I. Rabi's terse question, ``Who ordered that?'' quipped after the 1936 discovery of muon, eloquently states the essence of the family problem.   Let the matter representation be
\dis{
{\cal M}=\sum_{i=1}^{[N/2]} c _i [\, i\,]+\sum_{j=1}^{[N/2]} \bar{c} _j [ \,\bar{j}\,]-\bar{c} _{N/2} [ N/2]\delta_{N/2,\rm integer}\,.
}
where the elements of the integer set $\{c_i, \bar{c}_j \}$ do not have a common divisor.  For three standard families of Georgi-Glashow GUT \GG, Georgi required that all integers are 1 and found an SU(11) unification of three \GG~ families \cite{Georgi79}, which was the first exampe of UGUTF. Allowing possiblities of  $c_i>1$ and $ \bar c_j>1$ but requiring no common divisor for all of them is a non-repetition of a set, which is a reasonable requirement for a solution of the family problem. Then,  there are many possibilities \cite{FramptonPLB79}. If one requires renormalizable Yukawa couplings, in addition, with one irreducible type of representation containing the BEH doublet, an SU(9) model seems minimal \cite{FramptonPRL79}. However, at the GUT scale it is possible to have some Planck mass suppressed nonrenormalizable Yukawa couplings. Thus, constraining models by renormalizable couplings is not warranted at this stage. The minimal chiral choice plus vectorlike representations for UGUTF by the rules of (\ref{eq:NumberGGf}) is
\dis{
{\rm  SU(8)}\,:~  [3]\oplus[2]\oplus 9\,[\bar{1}]  \oplus n_{1}([1]\oplus [\bar{1}] ) \oplus n_{2}([2]\oplus [\bar{2}] )+\cdots.\label{eq:mini8}
}

\dis{
{\rm  SU(7)}\,:~  [3]\oplus 2\,[2]\oplus 8\,[\bar{1}]  \oplus n_{1}([1]\oplus [\bar{1}] ) \oplus n_{2}([2]\oplus [\bar{2}] )+\cdots.\label{eq:mini7}
}
where $n_i\ne 0$  for $i=1,2$ are needed for the BEH mechanism. For Eq. (\ref{eq:mini8}) the number of non-singlet chiral fields is 156, while for Eq. (\ref{eq:mini7}) the number of non-singlet chiral fields is 133. In this sense, Eq. (\ref{eq:mini7}) is the minimal model.

\subsection{Anti-SU($N$)'s and SO(32) heterotic string}

  By the rules of (\ref{eq:NumberGGf}), an SU(7) model with $[\bar3]\oplus  [2]\oplus  [\bar{1}] $ of Ref. \cite{Kim80} cannot be a theory of UGUTF   because family numbers in Eqs. (\ref{eq:NumberGGf})  do not allow funnily charged quarks and leptons. But it contains the key feature we explore in this paper. Spinor representations of SO($4n+2$) can be complex and the next possibility beyond  spinor {\bf 16} of SO(10)  is spinor {\bf 64} of SO(14) which contains the SU(7) model with $[\bar3]\oplus  [2]\oplus  [\bar{1}] $. Under the SU(7) antisymmetric-tensor notation, it is $ \Psi_{[\alpha\beta\gamma]}\oplus \Psi^{[\alpha\beta]}\oplus  \Psi_{[\alpha]} $ \cite{Kim80}. It was the first example of anti-SU($N$) where antisymmetric tensor BEH bosons, $\Phi^{[\alpha\beta]},\,\Phi_{[\alpha\beta]},\,\Phi^{[\alpha\beta\gamma]}$, and $\Phi_{[\alpha\beta\gamma]}$ are used to reduce the rank of the GUT gauge group. It allows two quark families and three lepton families, but it had a drawback due to the appearance of non-standard quarks and a doubly charged lepton. Note here that the rank $2n+1$ of SO($4n+2$) is reduced down to the rank $2n$  via the vacuum expectation values (VEVs) of  $\Phi^{[\alpha\beta]}$ and $\Phi_{[\alpha\beta]}$ representations of the SU($2n+1$)$\times$U(1) subgroup of  SO($4n+2$). 
 
Following the philosophy of Ref. \cite{Kim80}, we define SU($N$)$\times$U(1) subgroups of SO($2N$) by \antinD, \antinT, \antinQ, etc., which means that the spontaneous symmetry breaking of SO($2N$) takes the SU($N)\times$U(1)$_{2,3,4}$ subgroup route if $\Phi^{[AB]}$, $\Phi^{[ABC]}$, $\Phi^{[ABCD]}$ are used for the BEH bosons, respectively. The symmetry breaking of Ref. \cite{Kim80} was the intersection of \antiSD~and SU(7)$_{\rm anti3}$.   In this language, \flip~ is identical to \antiFD,
since there is no higher order anti-symmetric tensor field beyond $\Phi^{[AB]}$ for $N=5$.\footnote{In Ref. \cite{DKN84},  \flip ~was called anti-SU(5).}  In this vain, we can define SU$(N)_{\rm GG}$ as the symmetry breaking route of the BEH field $\Phi^{A}_B$ of SU($N)\times$U(1)  subgroup of SO($2N$).

It looks like that the SM families are only possible from the spinor of SO(10), which has been a reason for a GUT extension chain, containing the spinor {\bf 16} of SO(10),
\dis{
 {\rm SO(10)}\to {\rm E}_6\to {\rm E}_7\to {\rm E}_8,
  }
  or
  \dis{
{\rm SO(10)}\to {\rm SO(12)}\to {\rm SO(14)}\to {\rm SO(16)}\cdots
  }
Of course, SO($2N$) contains spinors, and  spinors for $N>5$ can produce {\bf 16}'s of SO(10).  In string theory,  gauge groups $\EE8$ and SO(32) are allowed only with their adjoint representations. This observation did not favor the SO(32) hetrotic string theory for the realistic purpose of obtaining {\bf 16} of SO(10), because the adjoint representation of SO(32) cannot produce any spinor representation in its subgroups. 

Because no non-SM charges have been found up to the TeV scale, the \antiFD~rather than  $[\bar3]\oplus  [2]\oplus  [\bar{1}] $ in SU(7) seems to contain some truth in it. But, for a UGUTF one should consider groups larger than SO(10). Here, we go beyond the  \antiFD. Then, we can use  anti-SU($N$)'s for a theory of UGUTF.
In particular, GUT symmetry breaking is easier in supersymmetric GUTs if we assign VEVs to $\Phi^{[45]}$ ($=\alpha_{45}$ of Eq. (\ref{eq:FundSUN})) and $\Phi_{[45]}$, \ie in \antinD.   In addition, the flipped SU(5) GUTs have been obtained from string compactification \cite{Ellis89, KimKyae07}. These string-derived anti- or flipped- SU(5)'s are assuming the symmetry breaking chain through the SO(10)  route assuming the appearance of the spinor {\bf 16}, but the SO(10)-spinor chain is not possible in Eq. (\ref{eq:NumberGGf}). Therefore, for UGUTFs  we follow the extension chain
\dis{
  {\rm SU(5)}  \to {\rm SU(6)} \to {\rm SU(7)} \to {\rm SU(8)} \to {\rm SU(9)}\left\{
 \begin{array}{l}  \to {\rm E_8}\\[0.3em]
 \to  \rm SO(18)\to  \cdots\to  SO(32).
  \end{array}
   \right. 
}
In these chains, we do not follow the groups allowing only spinor  representations.
So, anti-SU($N$) by anti-symmetric tensor fields \cite{Kim80} is the key in string compactification. The UGUTF of Eq. (\ref{eq:mini7}) may be obtained in this way. 

The SO(32) heterotic string is useful for phenomenology and UGUTF. Symmetry breaking of SO(32) through \antinD~is possible because {\bf 496} of SO(32) contains the following SU(16) representations 
\dis{
\Phi^A_B \oplus \Phi^{[AB]} \oplus \Phi_{[AB]}, ~~(A, B=1,2,\cdots,16),
}
whose dimensions are $(N^2-1), \frac{N(N-1)}{2}$, and $\frac{N(N-1)}{2}$, respectively, of its SU(16) subgroup. In the orbifold compactification of SO(32), it will be easy to realize the representation $\Phi^{[AB]}$ and $\Phi_{[AB]}$ even at level 1, and the key UGUTF  breaking, \ie the separation of color SU(3)$_c$ and weak SU(2)$_W$, to the SM is possible by $\langle \Phi^{[45]}\rangle$ and $\langle\Phi_{[45]}\rangle$ of Eq. (\ref{eq:FundSUN}). Because we allow only the SM fields, the fundamental representations $\Phi^{[A]}$ and  $\Phi_{[A]}$ are also used to reduce the rank further by the VEVs at the locations $f_6,\cdots,f_N$ of Eq. (\ref{eq:FundSUN}).

\section{Toward a theory of family uniflication in orbifold compactification}\label{section:Orb}

As noted in the previous section, a minimal UGUTF needs a GUT allowing SU(7) as a subgroup. Compactification of heterotic string frequently needs an anti-SU($N$) or \antinD. The $\EE8$ heterotic string allows the rank 8 SU(9) which is considered to be a subgroup of E$_8$. We will try to realize Eq. (\ref{eq:mini7}) from the string compactification. In addition,   no degeneracy between families is left below the compactification scale. This means that we will introduce all the needed Wilson lines. As we will see, it is more difficult to obtain multi-index tensor fields in the twisted sectors. In particular, the three-index tensor field $\Psi^{[ABC]}$ cannot be obtained in the twisted sectors. So, $\Psi^{ABC]}$, if they appear,  is required to come from $U$.  

\begin{table}[!h]
\begin{center}
\begin{tabular}{|c|c|c|c|}
 \hline &&&\\[-1.1em]
Order $N$ &   $\phi_s$  & $\phi_s^2$ &No. of fixed points\\[0.2em]
 \hline &&&\\[-1.1em]
  4 &   $\frac14(2~1~1)$ &$\frac{3}{8}$ & 16\\[0.2em] \hline &&&\\[-1.1em] 
  6-I  &$\frac16(2~1~1)$ &$\frac{1}{6}$& 3\\[0.2em] \hline &&&\\[-1.0em]
   6-II  & $\frac16(3~2~1)$&$\frac{14}{36}$ & 12\\[0.2em] \hline &&&\\[-1.1em] 
  8-I & $\frac18(3~2~1)$&$\frac{14}{64}$ & 4 \\[0.2em]\hline &&&\\[-1.1em]
 8-II  & $\frac18(4~3~1)$ &$\frac{26}{64}$& 8\\[0.2em] \hline &&&\\[-1.1em]
12-I  & $\frac1{12}(5~4~1)$&$\frac{42}{144}$ & 3\\[0.2em] \hline &&&\\[-1.1em] 
12-II  & $\frac1{12}(6~5~1)$&$\frac{62}{144}$ & 4 \\[0.2em]
\hline
\end{tabular}
 \caption{Number of fixed points of non-prime orbifolds.}\label{tab:lattices}
\end{center} 
\end{table}
In Table \ref{tab:lattices}, we list the number of fixed points for the non-prime orbifolds. For example, the cental number in $\Z_{6-II}$ and $\Z_{12-I}$ are 2 and 4, respectively, which mean that they have $\Z_{6/2}$ and $\Z_{12/4}$ symmetries, \ie both have the $\Z_3$ symmetry in the second torus. They correspond to the gauge group phase in the untwisted sector matter $P\cdot V=\frac{2}{6}$ and $\frac{4}{12}$, respectively, and the untwisted sector multiplicity is 3. This is the easiest way to obtain three families from the $U$ sector, \ie $3\,(\ten\oplus \fiveb)$. In this case, the three families are not distinguished, and there must be an $S_3$ discrete symmetry. It may be broken by Higgsing at a GUT scale. Here, we do not follow this line of argument  because there are too many possibilities. 

If $\Psi^{[ABC]}$ appears from $U$, its multiplicities can be 1, 2, 3, or 4. Always, $p\cdot V=\frac{1}{N}$ gives the multiplicity 1 except in $\Z_4$. In $\Z_{4}\,({\rm entries\,1})$ and $\Z_{8-I}\,({\rm entry\,2})$,  they give $\Z_4$ which leads to multiplicity 2. In $\Z_{6-I}\,({\rm entry\,2}), \Z_{6-II}\,({\rm entry\,2})$ and $\Z_{12-I}\,({\rm entry\,4})$, they give $\Z_3$ which leads to multiplicity 3. In $\Z_{4}\,({\rm entry\,2}), \Z_{6-II}\,({\rm entry\,3}),\Z_{8-II}\,({\rm entry\,4})$ and $\Z_{12-II}\,({\rm entry\,6})$, they give $\Z_2$ which leads to multiplicity 4. 

 \begin{table} [!h]
\begin{center}
\begin{tabular}{|c|c|c|c| c| }
 \hline  
$U_i$ &\,Number~of\,${\bf 10}\,\rm s$\,&\,Tensor~form\,&\,Chirality\,&\,$[p_{\rm spin}]~(p_{\rm spin}\cdot\phi_s)$\, \\[0.1em] \hline  
   ~$U_1\,(p\cdot V=\frac{5}{12})$~  & 1  &~$\Psi^{[ABC]}$~ &R &$[\oplus;+++]~~\left(\frac{+5}{12} \right)$  \\   
    $U_2\,(p\cdot V=\frac{4}{12})$ & 3 &   ~$\Psi^{[ABC]}$~&L &$[\ominus;++-]~~\left(\frac{+4}{12} \right)$    \\ 
    $U_3\,(p\cdot V=\frac{1}{12})$ &  1  &~$\Psi^{[ABC]}$~ &L  &$[\ominus;+-+]~~\left(\frac{+1}{12} \right)$   \\[0.1em]  
\hline
\end{tabular}
\end{center}
\caption{Number of $\ten$'s from \antiSD~in $\Z_{12-I}$, and chirality  for $P\cdot V=\phi_s\cdot s$ with $\phi_s=(\frac{5}{12},\frac{4}{12}, \frac{1}{12})$ and even number of minuses from $s=(\ominus~{\rm or}~\oplus;\pm,\pm,\pm)$. For example, $s=(\oplus;+++)$ gives chirality R for $P\cdot V=p_{\rm spin}\cdot\phi_s=\frac{5}{12}$. }
\label{tab:tenfromU}
\end{table}
 As an example, we show the $U$ sector multiplicity of $\Psi^{[ABC]}$ in $\Z_{12-I}$ in Table \ref{tab:tenfromU}. If it is the representation of an SU(8) GUT, then there are two $\ten$'s (due to Eq. (\ref{eq:NumberGGf})) from $U$.  If it is the representation of an SU(7) GUT, then there is a possibility of one $\ten$  from $U$. Thus, we choose \antiSD. The remaining  $\ten$'s of SU(5), \ie {\bf 21}'s of \antiSD, come from the twised sectors. However, it is not so easy to obtain two $\Psi^{[AB]}$'s of \antiSD~from $T$.
 
Summarizing the method to obtain three $\ten$'s from  \antiSD,
\dis{
\begin{array}{l}
\rm {\bf 1}.~ Matter~representation~ \Psi^{[ABC]}\,({A=1,2,\cdots,7})~ must~be ~present~in~ the~untwisted~ sector.\\ [0.2em]
\rm  {\bf 2}. ~Matter~\Psi^{[AB]}~ must~not ~appear~in~ the~untwisted~ sector. \\ [0.2em]
\rm {\bf 3}.~ Matter~\Psi^{[AB]}~ must~be ~present~in~ a~ twisted~ sector~with ~the ~chirality~that~ of~ \Psi^{[ABC]}. 
\end{array}\label{eq:conditions}
}
Matter in the untwisted sector $U_i$ occurs with $P\cdot V=\frac{N_i}{N}$.  For example,  $N_i=5,4,1$ for $\Z_{12-I}$ is shown in Table \ref{tab:lattices}. In addition, one matter $\Psi^{[ABC]}$ is allowed only in $U_1$ and $U_3$ in $\Z_{12-I}$. We require that it has the spinor form, by choosing an appropriate $V$,
\dis{
p\cdot V= \frac{1}{12},\,U_3 :&(\underline{----+++};+)(0^8)',    \\
p\cdot V  =\frac{5}{12},\,U_1 :&(\underline{----+ ++};+)(0^8)',
 \label{eq:Umatter}
}
where the underline means permutations. Here, the torus or untwisted sector are called  $U_3$ and $U_1$, respectively.
Their CTP conjugates also appear in $U$ as $(\underline{---++ ++};-)(0^8)'$.  Representations (\ref{eq:Umatter}) satisfies {\bf Condition 1}. To  satisfy {\bf Condition 2}, Wilson lines can be used if needed.  
  
{\bf Condition 3}  requires a full construction method. $\Psi^{[AB]}$ located in a twisted sector is the key part in this paper.  
In  the $\Z_{N}$ orbifold, multiplicities in the $k$-th twisted sector ${\cal P}_k$ need to be calculated, which is given by\footnote{$\tilde\chi(\theta^k,\theta^l)$ are presented in Ref. \cite{LNP696}.}
\dis{
{\cal P}_k   =\frac{1}{N}\sum_{l=0}^{[N/2]}
\tilde\chi(\theta^k,\theta^l)e^{i\,2\pi l\Theta_0},\label{eq:Multiplicity}
}
where $\Theta_0$ will be defined later.
The chirality is given by the first entry of $s$, denoted as L- or R- movers,  with the even number of total `$-$'s,  
\dis{
s=(s_0;\tilde{s})=(\ominus ~{\rm or}~\oplus\,;\pm,\pm,\pm).
}
 
\begin{table}[!h]
\begin{center} 
\begin{tabular}{|c|c|c cc|}
\hline
 &    &     &   {\rm Mult.} &\\
 & i    &   ${\cal P}_k (0)$ & ${\cal P}_k (\frac{\pi}{3})$&  ${\cal P}_k (\frac{2\pi}{3})$ \\[0.2em]
 \hline
    & 1       &3 &0 &0 \\
$\Z_6{\rm -I}$ & 2   &15 &0 &0  \\
 & 4    &   8&0&4  \\
  \hline 
     & 1       &12 &0 &0 \\
$\Z_6{\rm -II}$ & 2   &6 &0 &0  \\
 & 4   &   8&0&0 \\
  \hline  
  \end{tabular}
  \hskip 0.3cm
  \begin{tabular}{|c|c |ccc|}
\hline
 &    &  &{\rm Mult.} &\\
 & i  & ${\cal P}_k (0) $& ${\cal P}_k (\frac{\pi}{2})$&  ${\cal P}_k (\pi)$ \\[0.2em]
 \hline
    & 1    &4 &0 &0 \\
$\Z_8{\rm -I}$ & 2   &{\color{red} 10} &0 &0  \\
 & 3   &4 &0 &0 \\
 & 4  &6 &{ 3} &{  4}   \\
  \hline
    & 1   & 8 &0  & 0 \\
$\Z_8{\rm -II}$ &2 &{\color{red} 3}  & {\color{red} 1} &  {\color{red} 1}  \\
 & 3   & 8 & 0 & 0 \\
 & 4  &6&3 &{  4}   \\
  \hline
 \end{tabular}
 \hskip 0.3cm
\begin{tabular}{|c|c |cccc|}
\hline
 &  &&{\rm Mult.}&& \\
 & i  & ${\cal P}_k (0)$ & ${\cal P}_k (\frac{\pi}{3})$& ${\cal P}_k ( \frac{2\pi}{3} )$ & ${\cal P}_k (\pi)$   \\[0.3em]
 \hline
& 1   &3 &0&0&0\\
& 2  & 3 &0 &0&0 \\
$\Z_{12}{\rm -I}$ & 3  &{\color{red}2} &0&{\color{red}1} &0\\
& 4  &9 &0&0&6\\
& 5  &3&0&0&0\\
& 6  &{4}  &2  &{3}  &{2} \\
\hline
& 1  &4  &0 &0 &0 \\
& 2   &1  & 0 &0&0 \\
$\Z_{12}{\rm -II}$ & 3  & {\color{red} 8}  & 0&0 &0 \\
& 4 &5 &0&3&0\\
& 5 & 4 & 0& 0& 0\\
& 6  &{4}  &{2}  &{3}  &{2} \\
\hline
  \end{tabular}
\end{center}
\caption{Multiplicities in the $\Z_6, Z_8,$ and $\Z_{12}$ twisted sectors.}\label{tab:12Multiplicity}
\end{table}
 
In Table  \ref{tab:12Multiplicity}, multiplicities are presented for $\Z_{6}, \Z_{8}$ and $\Z_{12}$ orbifolds.  The multiplicities in $T_{N/4}$ are colored red. Matter $\Psi^{[AB]}$ can appear, if they do, only at $T_{N/4}$. From the red numbers in Table  \ref{tab:12Multiplicity}, we note that  $\Z_{8-II}$ and $\Z_{12-I}$ are the only allowed possibilities of obtaining multiplicity 2 at $T_{N/4}$, due to the large vacuum energy $2\tilde{c}_{N/4}=\frac{13}{8}$, as presented below in Eqs. (\ref{eq:Twist82}) and (\ref{eq:Twist121}). In other models, it is impossible to house $\Psi^{[AB]}$. Out of $\Z_{8-II}$ and $\Z_{12-I}$, we choose the simpler case  $\Z_{12-I}$ because we need to specify only one Wilson line. Note that in $\Z_{8-II}$ one has to specify two Wilson lines to specify the model completely without degeneracy. Then, at   $T_{N/4}$, we must obtain two $\Psi^{[AB]}$'s. This condition is very restrictive and may rule out the $\Z_{12-I}$ possibility. Luckily, we find a model, satisfying this condition.  

In the twisted sector, the masslessness conditions are satisfied for the phases contributed by the left- and right-movers \cite{KimKyae07},
\dis{
2N_L^j\hat{\phi}_j +(P+kV)\cdot V -\frac{k}{2}V^2 =2\tilde{c}_k, ~{\rm L~movers},\label{eq:oscillatorL}
}
\dis{
2N_R^j\hat{\phi}_j - \tilde{s}\cdot\phi_s +\frac{k}{2}\phi_s^2  =2 {c}_k, ~{\rm R~movers},\label{eq:oscillatorR}
}
where $j$ denotes the coordinate of the 6-dimensional compactified space running over $\{1,\bar{1}\},\{2,\bar{2}\},\{3,\bar{3}\}$, and $\hat{\phi}^j=\phi_{s}^j\cdot {\rm sign}(\tilde\phi^j)$ with  ${\rm sign}({\phi}^{\bar j})=-{\rm sign}(\tilde\phi^j) $. 
In calculating the multiplicities in Eq, (\ref{eq:Multiplicity}), we use the phase $\Theta_0$ with
$\Delta_k$,
\begin{eqnarray}
&\Theta_0  = \sum_j (N^j_L-N^j_R)\hat{\phi}^j -\frac{k}{2}(V_a^2-\phi_s^2)+(P+kV_a)\cdot V_a-(\tilde s +k\phi_s)\cdot\phi_s +{\rm integer}, \nonumber\\
&=-\tilde{s}\cdot\phi_s+\Delta_k,\label{eq:Phase}
\end{eqnarray}
\dis{
\Delta_k&= (P+kV_a)\cdot V_a-\frac{k}{2}(V_a^2-\phi_s^2)+\sum_j (N^j_L-N^j_R)\hat{\phi}^j\\
&\equiv\Delta_k^0+\Delta_k^N,\label{eq:Deltak}
}
where $V_a$ is $V$ distinguished by Wilson lines, and
\dis{
&\Delta_k^0 =P\cdot V_a+\frac{k}{2}(-V_a^2+\phi_s^2) , \\
&\Delta_k^N =\sum_j (N^j_L-N^j_R)\hat{\phi}^j.\label{eq:PhaseMore}
} 
We choose $0<\hat{\phi}^j\le 1$ mod integer and  oscillator contributions due to $(N_L-N_R)$ to the phase can be positive or negative, with $N_{L,R}\ge 0$. But each contribution to the vacuum energy $N_{L,R}^j\hat{\phi}^j$ is nonnegative. One oscillation contributes one number in $\phi_s$.  With the oscillator, the vacuum energy is shifted to
\dis{
&(P+kV_a)^2=-2\sum_j N_L^j \hat{\phi}^j +2\tilde{c}_k\\
&(p_{\rm vec}+k\phi_s)^2=-2\sum_j N_R^j \hat{\phi}^j +2 {c}_k .\label{eq:vacuumE}
}
In Eq. (\ref{eq:vacuumE}), instead of $p_{\rm vec}$ a four entry quantity $p_{\rm spin}$ of $\pm \frac12$'s with even number of $-$'s is possible, but we do not find any example with $p_{\rm spin}$.
The vacuum energy contributions in the twisted sectors in $\Z_{6,8,12}$ are given by\footnote{Typos of Appendix D of Ref. \cite{LNP696} are corrected here.}
\dis{
\Z_{6-I}: \left\{\begin{array}{ll}
2\tilde{c}_k:  &
 ~ \frac{3}{2}(k=1),~ \frac{4}{3}(k=2),~ \frac{3}{2}(k=3),\\[0.3em]
 2 {c}_k: &
  ~ \frac{1 }{ 2}(k=1),~ \frac{1}{3 }(k=2),~ \frac{1}{2}(k=3),  \end{array}\right.\label{eq:Twist61}
}
\dis{
\Z_{6-II}: \left\{\begin{array}{ll}
2\tilde{c}_k:  &
 ~ \frac{25}{18}(k=1),~ \frac{14}{9}(k=2),~ \frac{3}{2}(k=3),\\[0.3em]
 2 {c}_k: &
  ~ \frac{7}{ 18}(k=1),~ \frac{5}{9}(k=2),~ \frac{1}{2}(k=3),  \end{array}\right.\label{eq:Twist62}
}
\dis{
\Z_{8-I}: \left\{\begin{array}{ll}
2\tilde{c}_k:  &
 ~ \frac{47}{32}(k=1),~ \frac{11}{8}(k=2),~ \frac{47}{ 32}(k=3),~ \frac{3}{2}(k=4),\\[0.3em]
 2 {c}_k: &
  ~ \frac{15}{ 32}(k=1),~ \frac{3}{8}(k=2),~ \frac{15}{ 32}(k=1),~ \frac{1}{2}(k=3),  \end{array}\right.\label{eq:Twist81}
}
\dis{
\Z_{8-II}: \left\{\begin{array}{ll}
2\tilde{c}_k:  &
 ~ \frac{45}{32}(k=1),~ \frac{13}{8}(k=2),~ \frac{45}{ 32}(k=1),~ \frac{3}{2}(k=4),\\[0.3em]
 2 {c}_k: &
  ~ \frac{13}{32}(k=1),~ \frac{5}{8}(k=2),~ \frac{13}{ 32}(k=1),~ \frac{1}{2}(k=4).  \end{array}\right.\label{eq:Twist82}
}
\dis{
\Z_{12-I}: \left\{\begin{array}{ll}
2\tilde{c}_k:  &
 ~ \frac{210}{144}(k=1),~ \frac{216}{144}(k=2),~ \frac{234}{144}(k=3),~ \frac{192}{144}(k=4),~ \frac{210}{144}(k=5),~ \frac{216}{144}(k=6),\\[0.3em]
 2 {c}_k: &
  ~ \frac{11}{24}(k=1),~ \frac{1}{2}(k=2),~ \frac{5}{8}(k=3), ~\frac{1}{3}(k=4),~ \frac{11}{24}(k=5),~ \frac{1}{2}(k=6).  
 \end{array}\right.\label{eq:Twist121}
}
\dis{
\Z_{12-II}: \left\{\begin{array}{ll}
2\tilde{c}_k: &
 ~ \frac{103}{72}(k=1),~ \frac{31}{18}(k=2),~ \frac{11}{8}(k=3),~ \frac{14}{9}(k=4),~ \frac{103}{72}(k=5),~ \frac{3}{2}(k=6),\\[0.3em]
 2 {c}_k:&
  ~ \frac{31}{72}(k=1),~ \frac{13}{18}(k=2),~ \frac{3}{8}(k=3), ~\frac{5}{9}(k=4),~ \frac{31}{72}(k=5),~ \frac{1}{2}(k=6).  
 \end{array}\right.\label{eq:Twist122}
}
Note that $2\tilde{c}_k-2 {c}_k=1$ which is the required condition for ${\cal N}=1$ supersymmetry in the 4 dimensional spectra.

After $\Psi^{[ABC]}$ and $\Psi^{[AB]}$ are obtained, the number of families is fixed. Namely, the number of $\Psi_{[A]}$ of \antiSD~is fixed after fixing $\Psi^{[ABC]}$ and $\Psi^{[AB]}$. For $\Psi^{[ABC]}_R\oplus 2\Psi^{[AB]}_R$, there must be eight   $ \Psi_{[A]R}$'s.  Locating ${\five}_{\rm BEH}$ and ${\fiveb}_{\rm BEH}$ of \GG~in \antiSD~can be achieved in many ways.  

Not to allow any left-over degeneracy, one must assign all possible Wilson lines. For $\Z_3$ and $\Z_4$, one must specify three Wilson lines, $a_1=a_2, a_3=a_4, a_5=a_6$.  So, they have the most complicated Wilson line structures. For $\Z_{6-II}, \Z_{8-I}$ and $\Z_{8-II}$, one must specify two Wilson lines.  Specifying one Wilson line is enough  in $ \Z_{12-I}$, \ie only $a_3=a_4$ in the 2nd torus. 

The Wilson loop integral is basically the Bohm-Aharanov effect in the internal space of two-torus,
\dis{
\oint V^i\,dx_i=\frac12\oint \left(\vec{\nabla}\times \vec{V}\right)^{i}_{0,+,-}\epsilon^{ijk} \,dx_{[jk]}.\label{eq:BAflux}
}
If the {\bf B}-field (\ie $\vec{\nabla}\times \vec{V}$) at the orbifold singularity is present, the phase through $\Delta_0$ contributes in the multiplicity. For $\Z_{12-I}$, this is the case in $T_{1,2,4,5}$. The complication arises at the points with $3a_3=0$ mod. integer, \ie at $T_{3,6}$ \cite{KimKyae07},\footnote{$T_9$ contains the CTP conjugate states of $T_3$.}  
 where the Bohm-Aharanov phase has to be taken into account explicitly. If there is no {\bf B}-field at the orbifold singularity, there is no Bohm-Aharanov  phase, but then there for the (internal space) gauge symmetry we must require explicitly 
\dis{
(P+kV_0)\cdot a_3=0.\label{eq:WilsonCond}
 } 
This case applies to $ T_{3}^{0,+,-},T_6$,  and of course at $U$ also.  We distinguish $T_3$ by $0,+$ and $-$ because the phase  $\Delta_k^0$ of Eq. (\ref{eq:PhaseMore}) contains an extra $\frac{k}{2}$ factor.  Namely, Eq. (\ref{eq:WilsonCond}) is applied only at $U, T_3$ and $T_6$. We will comment more on this in Subsecs. $T_3$ and $T_6$.
  
\section{\antiSD~spectra}
\label{sec:SU7Model}

We calculate the SU(7) non-singlet spectra of \antiSD~in the $\Z_{12-I}$ orbifold.  
We choose the following model,
 \dis{ 
 V^a=\left\{\begin{array}{l} V_0=\left(\frac{-5}{12},\,\frac{-5}{12},\,\frac{-5}{ 12},\,\frac{-5}{12},\,\frac{-5}{12},\,\frac{-5}{12},\,\frac{-5}{12} ;\, \frac{+5}{12}\right)
\left(\frac{4}{12},\, \frac{4}{12},\, \frac{4}{12},\,\frac{4}{12},\, 0,\, \frac{4}{12},\, \frac{7}{12},\, \frac{ 3}{12}\right)',~V^2_0=\frac{ 338}{144},\\[0.2em]
a_3=a_4=(\frac{1}{3},\,\frac{1}{3},\,\frac{1}{3},\,\frac{1}{3},\,\frac{1}{3},\,\frac{1}{3},\,\frac{1}{3};\frac{1}{3}) (\frac{-1}{3},\,\frac{-1}{3},\,\frac{-1}{3},\,\frac{-1}{3},\,0,\,0,\, \frac{5}{3},-1)' .  \end{array}\right. \label{eq:Model}
}  
 Here, $a_3\,(=a_4)$ is chosen to allow and/or forbid some spectra,
and is composed of fractional numbers with the integer multiples of $\frac13$ because the second torus has the $\Z_3$ symmetry. Shifted lattices by Wilson lines are given by $V_+$ and $V_-$,
 \dis{ 
 \left\{\begin{array}{l}V_+=\left( \frac{-1}{12},\, \frac{-1}{12},\, \frac{-1}{12},\, \frac{-1}{12},\, \frac{-1}{12},\, \frac{-1}{12},\, \frac{-1}{12} ;  \,\frac{9}{12}\right)
\left(0,\, 0,\, 0,\, 0,\, 0,\, \frac{ 4}{12},\,\frac{27}{12},\, \frac{-9}{12} \right)',~V_+^2 =\frac{914 }{144},\\ [0.2em] 
V_-=   \left(\frac{-9}{12},\,\frac{-9}{12},\,\frac{-9}{12},\,\frac{-9}{12},\,\frac{-9}{12},\,\frac{-9}{12},\,\frac{-9}{12} ;\, \frac{+1}{12}\right) \left(
\frac{8}{12},\, \frac{8}{12},\, \frac{8}{12},\, \frac{8}{12},\, 0,\, \frac{4}{12},\,\frac{-13}{12},\, \frac{15}{12}\right)',~V_-^2=\frac{1234}{144}.\\[0.2em]
   \end{array}\right.
\label{eq:VPlusAndMinus}
}  


 We anticipated to achieve the key spectra needed for \antiSD\,  
\dis{
 \Psi^{[ABC]}_{U_3} + 2\Psi^{[AB]}_{T_3} +(?)\left(\Phi^{[AB]}_{T_3}+
 \Phi_{[AB],T_3}\right)+\cdots,\label{eq:TENinSU8}
}
where the sectors they appear are marked as subscripts. At this point, we do not fix how many vectorlike pairs appear in $T_3$.  The chiral representations, the candidates of fermion families, are represented by $\Psi$, and vectorlike reresentations, candidates for the BEH bosons, are represented by $\Phi$.

The orbifold conditions, toward a low energy 4 dimensional (4D) effective theory,  remove some weights of the original ten dimensional $\EE8$ weights. The remaining ones constitute the gauge multiplets and matter fields in the untwisted sector in the low energy 4D theory. Therefore, the weights in the $U$ sector must satisfy $P^2=2$ as in the original $\EE8$ weights. Orbifold conditions produce singularities. They are typically represented in three two-dimensional tori. A loop of string can be twisted around these singularities and define twisted sectors $T_k\,(k=1,2,\cdots,12)$. Twisting can introduce additional phases. Since $T_{12-k}$ provides the anti-particles of $T_k$, we consider only $T_k$ for $k=1,2,\cdots,6$. $T_6$ contains both particles and anti-particles. $T_6$, not affected by Wilson lines, is like an untwisted sector. It contains the antiparticles also as in $U$.
Since the Wilson lines can affect only in non-contractible loops as  the hidden sector Aharanov-Bohm effect, the Wilson lines can affect only around the singularities in the twisted sectors, but have no effects in the untwisted sector.  
 
  In this section we consider two twisted sectors,  $T_3 $ and $T_6$, explicitly. The other twisted sectors, $T_1, T_2, T_4, $ and $T_5$ will be listed in Appendix.

\subsection{Untwisted sector $U$}
In $U$, we find the following nonvanishing roots of SU(7)$\times$SU(4)$'$ 
\dis{
 {\rm E_8~gauge~ multiplet:}~~& P\cdot V =0{\rm~mod.~ integer}\\
{\rm SU(7)}&:\left\{\begin{array}{l} P=(\underline{+1\, -1\, 0\, 0\, 0\,  0\, 0};0)(0^8)'  \end{array}\right.\\[0.3em]
 {\rm E_8'~gauge~ multiplet:}~~& P\cdot V =0{\rm~mod.~ integer}\\
 {\rm SU(4)'} &: \left\{\begin{array}{l}  
 P=(0^8)( \underline{1\, -1\,0\,0 }~0~  0~  0~ 0  )'.
   \end{array} \right. 
}
In addition, there exists U(1)$^2\times$U(1)$'^{5}$ symmetry. 
The non-singlet SU(7) matter fields are
\dis{
 {\rm E_8~matter~ multiplet:}~~& P\cdot V =\frac{5}{12},{\rm~mod.~ integer}\\
{\rm SU(7)}&:\left\{\begin{array}{c}  
\left.
\begin{array}{l} 
P=(\underline{+++----};+)(0^8)'   \end{array} \right\}:  ~\Psi^{[ABC]}_R . \\[0.5em]
 \end{array}\right.
 }
$\Psi^{[ABC]}$ contains one {\bf 10} of \GG~family, which belongs to the first family, and is R-handed as shown in Table \ref{tab:tenfromU}. 
It is simple to find the E$_8'$ hidden sector matter in $U$,
 \dis{
  {\rm E_8'~matter~ multiplet:}~~&\\
{\rm SU(4)'}&:\left\{\begin{array}{l}  
   P=(0^8)(\underline{+---};+---)',~ P\cdot V =\frac{1}{12} \end{array}\right\} :~  \four'_L , 
 }
 where $P\cdot a_3=0$ excludes the cases of $P=(0^8)(\underline{+---};+++-)'\,(P\cdot V= 0)$ and $P=(0^8)(\underline{++--};-++-)'\,(P\cdot V= \frac{4}{12})$.

\subsection{Twisted sectors $T$}
In the twisted sectors, we list only SU(7) or SU(4)$'$ non-singlets. Wilson lines distinguish three fixed points in the second torus, and the shift vectors we consider at $T_k$ are   split into three cases 
\dis{
T_k^{0,+,-}: ~kV_a=\left\{\begin{array}{l}kV\equiv kV_0\\[0.2em]
k(V+a_3)\equiv kV_+\\[0.2em]
k(V-a_3)\equiv kV_-,\end{array}\right.
}
where $V_+$ and $V_-$ are given in Eq. (\ref{eq:VPlusAndMinus}).
Because $3a_3=0$ mod. integer and due to Eq. (\ref{eq:Deltak}), $T_{ 6 }$ sector  is not distingushed by the Wilson lines but $T_{3,9}$ are distinguished. 

We select only the even lattices shifted from the untwisted lattices. They form even numbers for the sum of entries of each elements of $P$. 

 To obtain non-trivial number of families, we need two index tensor fields, $\Psi^{[AB]}$ and/or $\Psi_{[AB]}$. This possibility arises only in $T_3$ because at $T_3$ there appear fractional number with integer times $\frac14$. In other twisted sectors, the entries are not multiples of $\frac14$ in which case we cannot fulfil the masslessness condition with $2\tilde{c}_k$ given below.
 
In the $k$-th twisted sector, the masslessness condition to raise the tachyonic vacuum energy to zero is  
\dis{
(P+kV_a)^2=2\tilde{c}_k -(2\sum_j N_L^j \hat{\phi}^j ),\label{eq:MasslessConV}
}
\dis{
({\it p}+k\phi_s)^2=2{c}_k -(2\sum_j N_R^j \hat{\phi}^j ),\label{eq:MasslessConS}
}
where $2\tilde{c}_k$ and $2 {c}_k$ are given in Eq.   (\ref{eq:Twist121}), and the brackets must be taken into account when oscillators contribute.
When the conditions (\ref{eq:MasslessConV},\ref{eq:MasslessConS}) are satisfied, we obtain the SUSY spectra for which the chirality and multiplicity are calculated from $\Theta_0$ in the $k$-th twisted sector,

\begin{eqnarray}
&\Theta_0   =-\tilde{s}\cdot\phi_s+ P\cdot V_a+\Delta_k^0+\Delta_k^N+\left( {\it p}\cdot \phi_{s}+\delta_k^N\right) ,\label{eq:PhaseA}
\end{eqnarray}
where
\dis{
&\Delta_k^0 = \frac{k}{2}(\phi_s^2-V_a^2) , \\
&\Delta_k^N =2\sum_j  N^j_L\,\hat{\phi}^j , \\
&\delta_k^N =-2\sum_j   N^j_R\,\hat{\phi}^j.\label{eq:DeltaNkA}
}
We choose $0<\hat{\phi}^j\le 1$ mod integer and  oscillator contributions due to $(N_L-N_R)$ can be in principle positive or negative. As an example, consider the $T_3$ sector for the $N_R^j$ contribution.
Here, $3\phi_s=(\frac54,\frac44,\frac14)$, needing ${\it p}=(-1,-1,0)$. So, $({\it p}+3\phi_s)^2=\frac18$, needing the $N_R$ contribution $\frac48=2\cdot \frac{3}{12}$ to make up $2c_3=\frac58$  \cite{KimKyae07}. Thus, the R-handed oscillator contribution is $\frac{6}{12}=2\cdot\frac{3}{12}$. Namely, $\delta^3$ is  $\frac{+3}{12}$ which is included in the tables.

We will select only the even lattices shifted from the untwisted lattices. They form even numbers if the entries of each elements of $P$ are added. Because we consider $\EE8$, the product of $E_8$ and $E_8'$ parts  must be even. They need not be even separately. But, there is a distinction in ${\rm even}\times{\rm even}$ lattice and  ${\rm odd}\times{\rm odd}$ lattice. In the former case, E$_8$ and E$_8'$ gauge quatum numbers do not change, but in the latter case we change the signs of the quantum numbers. In the table captions, we take into account this fact.
   
\subsubsection{Twisted sector $T_3 \,( \delta^3 =\frac{+3}{12})$}
In the multiplicity calculation in $\Theta_0$, there is a factor $\frac12$ between the lattice shifts by Wilson lines. Even though the Wilson lines cannot distinguish the fixed points, we consider $V_+$ and $V_-$ also as if Wilson lines distinguish fixed points.  
 
\begin{itemize}
\item Two index spinor form for $V^3_{0}$:
 the spinor form gives  $(P+3V_0)^2=\frac{13}{8}$  but $(P+3V_0)\cdot a_3\ne 0$, and there is no allowed states.
  
\item  Two index vector form  for $V_0^3 $:
For  a vector form of $P$,  
 \dis{
&3V_0=\left(\frac{-5}{4},\,\frac{-5}{4},\,\frac{-5}{4},\,\frac{-5}{4},\,\frac{-5}{4},\,\frac{-5}{4},\,\frac{-5}{4} ;\, \frac{+5}{4}\right)
\left(\frac{4}{4},\, \frac{4}{4},\, \frac{4}{4},\,\frac{4}{4},\, 0,\, \frac{4}{4},\, \frac{7}{4},\, \frac{ 3}{4}\right)',~V^2_0=\frac{ 338}{144}\\ 
&P =(\underline{  2, 2,1,1,1,1,1 };-1)(-1,-1,-1,-1,0,-1,-2,-1)',  ~P\cdot V_0= \frac{ -87 }{12}\\
&P+3V_0  =\left(\underline{\frac{ 3}4\, \frac{ 3}4\, \frac{ -1}4\, \frac{-1}4\, \frac{ -1}4\, \frac{ -1}4\, \frac{ -1}4}; \frac{ 1}4 \right)\left(0^6\, \frac{-1}{4}\, \frac{ -1}4  \right)',~ (P+3V_0)\cdot a_3= 0,\\
&3\phi_s=\left(\frac{5}{4},\,\frac{4}{4},\,\frac{1}{4} \right) ,~p_{\rm vec} =(-1,-1,0),~p_{\rm vec}\cdot\phi_s =\frac{-9}{12},~p_{\rm vec} +3\phi_s  =\left( \frac{1}{4},0, \frac{1}4  \right).
} 
Massless states are shown in Table \ref{tab:T3ZeroVec}.
Since the phase $\frac{21}{48}-\frac{169}{48}$ is an even integer times $\frac{1}{24}$, we choose the vector form $-p_{\rm vec}\cdot\phi_s =-[\frac{5}{12}\times (-1) +\frac{4}{12} \times (-1)+\frac{1}{12}\times (0)]=\frac{18}{12}$,   instead of the spinor form $p_{\rm spin}$. So, we used $-p_{\rm vec}\cdot\phi_s =\frac{+9}{12}$ in the table. 
  \begin{table}[!h]
\begin{center}
\begin{tabular}{|cc|c|c|ccc|c|c|c| }
 \hline &&&&&&&  \\[-1.15em]
 Chirality  &   $\tilde s$& $-\tilde{s}\cdot\phi_s$& ${\rm -}p_{\rm vec}\cdot\phi_s ,~P\cdot V_0$& $(3/2)\phi_s^2$,&$  -(3/2) V_0^2 $,& $\Delta_{3}^N[\delta^3 ] $   &$\Theta_0\,( {\cal P}_3^N)$    \\[0.15em]
 \hline &&&&&&&   \\[-1.15em]
$\ominus=L$&  $(---)$  &  $\frac{+5}{12}$ & $ ~~\frac{+9}{12},~~~~~~\frac{-3}{12}$ &$\frac{21}{48} $  &$\frac{-169}{48}$ & $0  [\frac{+3}{12}],0 [\frac{-3}{12}]   $& 
$\frac{+ 1}{12}\,(0),\, \frac{+7}{12}\,(0) $   \\ [0.1em] \hline &&&&&&& \\[-1.25em] 
$\ominus=L$ &   
$\color{red} (-++)$&  $0$ &$ ~~\frac{+9}{12},~~~~~~\frac{-3}{12}$&$\frac{21}{48}$&$\frac{-169}{48}$& 
$0 [\frac{+3}{12}],0 [\frac{-3}{12}]  $ &$\frac{+8}{12}\,(3),\, \frac{+2}{12}\,(2)  $  \\[0.15em] 
  \hline  &&&&&&&  \\[-1.25em]
  $\ominus=L$&  $ (+-+)$  & $\frac{-1}{12}$ &$ ~~\frac{+9}{12},~~~~~~\frac{-3}{12}$ &$\frac{21}{48}$&$\frac{-169}{48}$& 
 $0 [\frac{+3}{12}],0 [\frac{-3}{12}]   $ &  $\frac{ +7}{12}\,(0),\, \frac{+1}{12}\,(0)  $  \\[0.15em]  \hline &&&&&&&  \\[-1.25em]
 $\ominus=L$&  $ \color{red}(++-)$ &$\frac{-4}{12}$&$ ~~\frac{+9}{12},~~~~~~\frac{-3}{12}$ &$\frac{21}{48}$&$\frac{-169}{48}$&$0 [\frac{+3}{12}],0 [\frac{-3}{12}]   $ &$\frac{+4}{12}\,(3),\, \frac{-2}{12}\,(2) $   \\ [0.1em]  \hline &&&&&&&  \\[-1.25em]
$\oplus=R$ &  $(+++)$&  $\frac{-5}{12}$ &$ ~~\frac{+9}{12},~~~~~~\frac{-3}{12}$ &$\frac{21}{48}$&$\frac{-169}{48}$& $0 [\frac{+3}{12}],0 [\frac{-3}{12}]  $&$\frac{+3}{12}(0),\, \frac{-3}{12}\,(0)  $    \\[0.1em]   \hline &&&&& &&  \\[-1.25em]
  $\oplus=R$ &   $ \color{red}(+--)$& $0$ &$ ~~\frac{+9}{12},~~~~~~\frac{-3}{12}$ &$\frac{21}{48}$&$\frac{-169}{48}$ &
$0 [\frac{+3}{12}],0 [\frac{-3}{12}]   $ &$\frac{+8}{12}(3),\, \frac{+2}{12}\,(2)  $   \\[0.1em] \hline &&&&&&&  \\[-1.25em]
$\oplus=R$ &   $ (-+-)$& $\frac{+1}{12}$ &$ ~~\frac{+9}{12},~~~~~~\frac{-3}{12}$ &$\frac{21}{48}$&$\frac{-169}{48}$ &
 $0  [\frac{+3}{12}],0 [\frac{-3}{12}]  $&$\frac{+9}{12}\,(0),\, \frac{+3}{12}\,(0)  $   \\[0.1em] \hline &&&&&&& \\[-1.25em]
$\oplus=R$&  $ \color{red}(--+)$  &  $\frac{+4}{12}$ &$  \color{red}~~\frac{+9}{12},~~~~~~\frac{-3}{12}$ &$\frac{21}{48}$&$\frac{-169}{48} $& $0 [\frac{+8}{12}],0 [\frac{-3}{12}]  $&  $\frac{+12}{12}\,(4),\, \frac{+6}{12}\,(2)  $   \\[0.15em] 
\hline
\end{tabular}
\end{center}
\caption{Two index vector for $V^3_{0}$: Chiralities (in the first column) and multiplicities (in the last column) of $\Phi^{[AB]}$ in the $T_3$ sector of $\Z_{12-I}$ for $N=12$. $\phi_s=(\frac{5}{12},\frac{4}{12},\frac{1}{12}),\,(3/2)\phi_s=(\frac{5}{8},\frac{4}{8},\frac{1}{8})$ and $12(P+3V)\cdot a_3=0$. In the last column, $\delta^k_R=\frac{3}{12}$ is added. Multiplicities of the masslessness states are given by the phase  $\Theta_0$. The allowed chiralities  are colored red, ${\color{red} \Psi^{[AB]}_R}+10(\Phi^{[AB]}_L+\Phi^{[AB]}_R)$.} \label{tab:T3ZeroVec}
\end{table}

\end{itemize}

\begin{itemize}

\item Two index spinor form for $V_+^3$:
 the spinor form gives  $(P+3V_0)^2=\frac{13}{8}$  but $(P+3V_0)\cdot a_3\ne 0$, and there is no allowed states.
  
\item  Two index vector form  for $V_+^3 $:
For  a vector form of $P$,  
 \dis{
 &3V_+=\left( \frac{-1}{4},\, \frac{-1}{4},\, \frac{-1}{4},\, \frac{-1}{4},\, \frac{-1}{4},\, \frac{-1}{4},\, \frac{-1}{4} ;  \,\frac{9}{4}\right)
\left(0,\, 0,\, 0,\, 0,\, 0,\, \frac{ 4}{4},\,\frac{27}{4},\, \frac{-9}{4} \right)',~V_+^2 =\frac{914 }{144}\\  
&P =(\underline{  2, 2,1,1,1,1,1 };-1)(-1,-1,-1,-1,0,-1,-7, 2)',  ~P\cdot V_+= \frac{ -219 }{12}=\frac{+9}{12}\\
&P+3V_0  =\left(\underline{\frac{ 3}4\, \frac{ 3}4\, \frac{ -1}4\, \frac{-1}4\, \frac{ -1}4\, \frac{ -1}4\, \frac{ -1}4}; \frac{ 1}4 \right)\left(0^6\, \frac{-1}{4}\, \frac{ -1}4  \right)',~ (P+3V_0)\cdot a_3= 0,\\
&3\phi_s=\left(\frac{5}{4},\,\frac{4}{4},\,\frac{1}{4} \right) ,~p_{\rm vec} =(-1,-1,0),~p_{\rm vec}\cdot\phi_s =\frac{-9}{12},~p_{\rm vec} +3\phi_s  =\left( \frac{1}{4},0, \frac{1}4  \right).
} 
Massless states are shown in Table \ref{tab:T3PlusVec}, which are exactly the same as those of Table \ref{tab:T3ZeroVec}. 
  \begin{table}[!h]
\begin{center}
\begin{tabular}{|cc|c|c|ccc|c|c| }
 \hline &&&&&&& \\[-1.15em]
 Chirality  &   $\tilde s$& $-\tilde{s}\cdot\phi_s$& ${\rm -}p_{\rm vec}\cdot\phi_s ,~P\cdot V_+$& $(3/2)\phi_s^2$,&$  -(3/2) V_+^2 $,& $\Delta_{3}^N[\delta^3 ] $   &$\Theta_0\,( {\cal P}_3^N)$    \\[0.15em]
 \hline &&&&&&&  \\[-1.15em]
$\ominus=L$&  $ (---)$  &  $\frac{+5}{12}$ & $ ~~\frac{+9}{12},~~~~~~\frac{-3}{12}$ &$\frac{21}{48} $  &$\frac{-169}{48}$ & $0  [\frac{+3}{12}],0 [\frac{-3}{12}]   $& 
$\frac{+ 1}{12}\,(0),\, \frac{+7}{12}\,(0) $  \\ [0.1em] \hline &&&&&&& \\[-1.25em] 
$\ominus=L$ &   
$\color{red} (-++)$&  $0$ &$ ~~\frac{+9}{12},~~~~~~\frac{-3}{12}$&$\frac{21}{48}$&$(\frac{-1}{12})\frac{-457}{48}$& 
$0 [\frac{+3}{12}],0 [\frac{-3}{12}]  $ &$\frac{+8}{12}\,(3),\, \frac{+2}{12}\,(2)  $  \\[0.15em] 
  \hline  &&&&&&&  \\[-1.25em]
  $\ominus=L$&  $ (+-+)$  & $\frac{-1}{12}$ &$ ~~\frac{+9}{12},~~~~~~\frac{-3}{12}$ &$\frac{21}{48}$&$\frac{-169}{48}$& 
 $0 [\frac{+3}{12}],0 [\frac{-3}{12}]   $ &  $\frac{ +7}{12}\,(0),\, \frac{+1}{12}\,(0)  $  \\[0.15em]  \hline &&&&&&&  \\[-1.25em]
 $\ominus=L$&  $\color{red} (++-)$ &$\frac{-4}{12}$&$ ~~\frac{+9}{12},~~~~~~\frac{-3}{12}$ &$\frac{21}{48}$&$\frac{-169}{48}$&$0 [\frac{+3}{12}],0 [\frac{-3}{12}]   $ &$\frac{+4}{12}\,(3),\, \frac{-2}{12}\,(2) $   \\ [0.1em]  \hline &&&&&&& \\[-1.25em]
$\oplus=R$ &  $(+++)$&  $\frac{-5}{12}$ &$ ~~\frac{+9}{12},~~~~~~\frac{-3}{12}$ &$\frac{21}{48}$&$\frac{-169}{48}$& $0 [\frac{+3}{12}],0 [\frac{-3}{12}]  $&$\frac{+3}{12}(0),\, \frac{-3}{12}\,(0)  $   \\[0.1em]   \hline &&&&& &&  \\[-1.25em]
  $\oplus=R$ &   $\color{red} (+--)$& $0$ &$ ~~\frac{+9}{12},~~~~~~\frac{-3}{12}$ &$\frac{21}{48}$&$\frac{-169}{48}$ &
$0 [\frac{+3}{12}],0 [\frac{-3}{12}]   $ &$\frac{+8}{12}(3),\, \frac{+2}{12}\,(2)  $  \\[0.1em] \hline &&&&&&& \\[-1.25em]
$\oplus=R$ &   $  (-+-)$& $\frac{+1}{12}$ &$ ~~\frac{+9}{12},~~~~~~\frac{-3}{12}$ &$\frac{21}{48}$&$\frac{-169}{48}$ &
 $0  [\frac{+3}{12}],0 [\frac{-3}{12}]  $&$\frac{+9}{12}\,(0),\, \frac{+3}{12}\,(0)  $  \\[0.1em] \hline &&&&&&&  \\[-1.25em]
$\oplus=R$&  $\color{red} (--+)$  &  $\frac{+4}{12}$ &$\color{red} ~~\frac{+9}{12},~~~~~~\frac{-3}{12}$ &$\frac{21}{48}$&$\frac{-169}{48} $& $0 [\frac{+8}{12}],0 [\frac{-3}{12}]  $&  $\frac{+12}{12}\,(4),\, \frac{+6}{12}\,(2)  $   \\[0.15em] 
\hline
\end{tabular}
\end{center}
\caption{Two index vector for $V_+^3$: The entries are the same as Table \ref{tab:T3ZeroVec}, and we obtain ${\color{red} \Psi^{[AB]}_R}+10(\Phi^{[AB]}_L+\Phi^{[AB]}_R)$.} \label{tab:T3PlusVec}
\end{table}

\end{itemize}

\begin{itemize}

\item Two index spinor form for $V_-^3$:
 the spinor form gives  $(P+3V_0)^2=\frac{13}{8}$  but $(P+3V_0)\cdot a_3\ne 0$, and there is no allowed states.
  
\item  Two index vector form  for $V_-^3 $:
For  a vector form of $P$,  
 \dis{
&3V_-=   \left(\frac{-9}{4},\,\frac{-9}{4},\,\frac{-9}{4},\,\frac{-9}{4},\,\frac{-9}{4},\,\frac{-9}{4},\,\frac{-9}{4} ;\, \frac{+1}{4}\right) \left(
\frac{8}{4},\, \frac{8}{4},\, \frac{8}{4},\, \frac{8}{4},\, 0,\, \frac{4}{4},\,\frac{-13}{4},\, \frac{15}{4}\right)',~V_-^2=\frac{1234}{144}\\
&P =(\underline{  2, 2,1,1,1,1,1 };-1)(-2,-2,-2,-2,0,-1,3,-4)',  ~P\cdot V_-= \frac{ -100 }{12}=\frac{-4}{12},\\
&P+3V_-  =\left(\underline{\frac{ 3}4\, \frac{ 3}4\, \frac{ -1}4\, \frac{-1}4\, \frac{ -1}4\, \frac{ -1}4\, \frac{ -1}4}; \frac{ 1}4 \right)\left(0^6\, \frac{-1}{4}\, \frac{ -1}4  \right)',~ (P+3V_0)\cdot a_3= 0,\\
&3\phi_s=\left(\frac{5}{4},\,\frac{4}{4},\,\frac{1}{4} \right) ,~p_{\rm vec} =(-1,-1,0),~p_{\rm vec}\cdot\phi_s =\frac{-9}{12},~p_{\rm vec} +3\phi_s  =\left( \frac{1}{4},0, \frac{1}4  \right).
} 
Massless states are shown in Table \ref{tab:T3MinusVec}. 
  \begin{table}[!h]
\begin{center}
\begin{tabular}{|cc|c|c|ccc|c|c| }
 \hline &&&&&&& \\[-1.15em]
 Chirality  &   $\tilde s$& $-\tilde{s}\cdot\phi_s$& ${\rm -}p_{\rm vec}\cdot\phi_s ,~P\cdot V_-$& $(3/2)\phi_s^2$,&$  -(3/2) V_-^2 $,& $\Delta_{3}^N[\delta^3 ] ,\Delta_{3}^N[-\delta^3 ]$   &$\Theta_0\,( {\cal P}_3^N)$    \\[0.15em]
 \hline &&&&&&&  \\[-1.15em]
$\ominus=L$&  $(---)$  &  $\frac{+5}{12}$ & $ ~~\frac{+9}{12},~~~~~~\frac{-4}{12}$ &$\frac{21}{48} $  &$\frac{-617}{48}$ & $0  [\frac{+3}{12}],0 [\frac{-3}{12}]   $& 
$\frac{+8}{12}\,(3),\, \frac{+2}{12}\,(2)$  \\ [0.1em] \hline &&&&&&& \\[-1.25em] 
$\ominus=L$ &   
$ (-++)$&  $0$ &$ ~~\frac{+9}{12},~~~~~~\frac{-4}{12}$&$\frac{21}{48}$&$ \frac{-617}{48}$& 
$0 [\frac{+3}{12}],0 [\frac{-3}{12}]  $ &$\frac{+3}{12}\,(0),\, \frac{-3}{12}\,(0)  $  \\[0.15em] 
  \hline  &&&&&&&  \\[-1.25em]
  $\ominus=L$&  $ (+-+)$  & $\frac{-1}{12}$ &$ ~~\frac{+9}{12},~~~~~~\frac{-4}{12}$ &$\frac{21}{48}$&$\frac{-617}{48}$& 
 $0 [\frac{+3}{12}],0 [\frac{-3}{12}]   $ &  $\frac{+2}{12}\,(2),\, \frac{-4}{12}\,(3)$  \\[0.15em]  \hline &&&&&&&  \\[-1.25em]
 $\ominus=L$&  $ (++-)$ &$\frac{-4}{12}$&$ ~~\frac{+9}{12},~~~~~~\frac{-4}{12}$ &$\frac{21}{48}$&$\frac{-617}{48}$&$0 [\frac{+3}{12}],0 [\frac{-3}{12}]   $ &$\frac{-1}{12}\,(0),\, \frac{-7}{12}\,(0)$   \\ [0.1em]  \hline &&&&&&& \\[-1.25em]
$\oplus=R$ &  $(+++)$&  $\frac{-5}{12}$ &$ ~~\frac{+9}{12},~~~~~~\frac{-4}{12}$ &$\frac{21}{48}$&$\frac{-617}{48}$& $0 [\frac{+3}{12}],0 [\frac{-3}{12}]  $&$\frac{-2}{12}(2),\, \frac{-8}{12}\,(3)  $   \\[0.1em]   \hline &&&&& &&  \\[-1.25em]
  $\oplus=R$ &   $ (+--)$& $0$ &$ ~~\frac{+9}{12},~~~~~~\frac{-4}{12}$ &$\frac{21}{48}$&$\frac{-617}{48}$ &
$0 [\frac{+3}{12}],0 [\frac{-3}{12}]   $ &$\frac{+3}{12}\,(0),\, \frac{-3}{12}\,(0) $  \\[0.1em] \hline &&&&&&& \\[-1.25em]
$\oplus=R$ &   $ (-+-)$& $\frac{+1}{12}$ &$ ~~\frac{+9}{12},~~~~~~\frac{-4}{12}$ &$\frac{21}{48}$&$\frac{-617}{48}$ &
 $0  [\frac{+3}{12}],0 [\frac{-3}{12}]  $&$\frac{+4}{12}\,(3),\, \frac{-2}{12}\,(2)$  \\[0.1em] \hline &&&&&&&  \\[-1.25em]
$\oplus=R$&  $ (--+)$  &  $\frac{+4}{12}$ &$ ~~\frac{+9}{12},~~~~~~\frac{-4}{12}$ &$\frac{21}{48}$&$\frac{-617}{48} $& $0 [\frac{+8}{12}],0 [\frac{-3}{12}]  $&  $\frac{+7}{12}\,(0),\, \frac{+1}{12}\,(0)$   \\[0.15em] 
\hline
\end{tabular}
\end{center}
\caption{ Two index vector for $V_-^3$: Chiralities (in the first column) and multiplicities are $ 10(\Phi^{[AB]}_L+\Phi^{[AB]}_R)$.} \label{tab:T3MinusVec}
\end{table}

\end{itemize}

\noindent The chiral spectrum we obtained for the two index tensors in $T_3$ is
\dis{
  T_3\,:~ 2\,\Psi^{[AB]}_{R,T_3^0}   .\label{eq:Tenfields}
}
These make up three chiral families together with $\Psi^{[ABC]}_R$ from $U$. The number in (\ref{eq:Tenfields}) is the same as the one if we treat $T_3$ with multiplicity 2. This multiplicity is because $3V$ is a $\Z_4$ twist which has two fixed points in a two-dimensional torus. Since there is no Wilson line,  we could have treated only $3V$ with multiplicity 2 of $\Z_4$. The multiplicity 2 is accounted by $T_3^0$ and $T_3^+$. But,  $T_3^-$ produce additional vectorlike pairs, which must be fictitious. We may consider the spectra in Table \ref{tab:T3MinusVec} are fictitious. In the remainder of the paper, we will not consider $V_+^3$ and $V_-^3$. We consider only  $V_0^3$ and take into account the multiplicity 2 of $T_3$. 

Now let us proceed to consider one index tensors in $T_3$. In the final result, we will multiply the overall multiplicity 2 as commented above.

\begin{itemize}

\item  One index spinor form  for $V^3_{0}$:

 \dis{
&3V_0=\left(\frac{-5}{4},\,\frac{-5}{4},\,\frac{-5}{4},\,\frac{-5}{4},\,\frac{-5}{4},\,\frac{-5}{4},\,\frac{-5}{4} ;\, \frac{+5}{4}\right)
\left(\frac{4}{4},\, \frac{4}{4},\, \frac{4}{4},\,\frac{4}{4},\, 0,\, \frac{4}{4},\, \frac{7}{4},\, \frac{ 3}{4}\right)',~V^2_0=\frac{ 338}{144}\\
&P =\left(\underline{\frac{1}{2}, \frac{3}{2}, \frac{3}{2}, \frac{3}{2}, \frac{3}{2}, \frac{3}{2}, \frac{3}{2}}; \frac{-3}{2}\right)(-1,-1,-1,-1,0,-1,-2,-1)',~P\cdot V_0=\frac{-92}{12}\\
&P+3V_0  =\left(\underline{\frac{- 3}4\, \frac{ 1}4\, \frac{ 1}4\, \frac{ 1}4\, \frac{ 1}4\, \frac{ 1}4\, \frac{ 1}4}; \frac{ -1}4 \right)\left(0^6\, \frac{-1}{4}\, \frac{-1}4  \right)',~ (P+3V_0)\cdot a_3= 0,
} 
which make up $\frac98$. The oscillator contributions of $2\frac{3}{12}$ are needed to satisfy the masslessness condition. Chiralities and multiplicities are tabulated in  Table \ref{tab:T3ZeroOneInSpinor}.
  \begin{table}[!h]
\begin{center}
\begin{tabular}{|cc|c|c|ccc|c|c|}
 \hline &&&&&&&   \\[-1.15em]
 Chirality  &   $\tilde s$& $-\tilde{s}\cdot\phi_s$& ${\rm -}p_{\rm vec}\cdot\phi_s ,~P\cdot V_0$& $(3/2)\phi_s^2$,&$  -(3/2) V_0^2 $,& $\pm \Delta_{3}^N[\pm \delta^3 ] $   &$\Theta_0\,( {\cal P}_3^N)$     \\[0.15em]
 \hline &&&&&&&   \\[-1.15em]
$\ominus=L$&  $\color{red}(---)$  &  $\frac{+5}{12}$ & $ \color{red} ~~\frac{+9}{12},~~~~~~\frac{+4}{12}$ &$\frac{21}{48} $  &$\frac{-169}{48}$ & $\frac{\pm 3}{12}[\frac{\pm 3}{12}] $& 
$\frac{+12}{12}(4), \frac{+6}{12}(2), \frac{+6}{12}(2), \frac{0}{12}(4)$  \\ [0.1em] \hline &&&&&&& \\[-1.15em] 
$\ominus=L$ &   
$  (-++)$&  $0$ &$ ~~\frac{+9}{12},~~~~~~\frac{+4}{12}$&$\frac{21}{48}$&$\frac{-169}{48}$& 
$ \frac{\pm 3}{12}[\frac{\pm 3}{12}] $ &$ \frac{+7}{12}(0), \frac{+1}{12}(0), \frac{+1}{12}(0), \frac{-5}{12}(0) $  \\[0.15em] 
  \hline  &&&&&&&  \\[-1.15em]
  $\ominus=L$&  $\color{red} (+-+)$  & $\frac{-1}{12}$ &$ \color{red} ~~\frac{+9}{12},~~~~~~\frac{+4}{12}$ &$\frac{21}{48}$&$\frac{-165}{48}$& 
 $\frac{\pm 3}{12}[\frac{\pm 3}{12}] $ &  $ \frac{+6}{12}(2), \frac{0}{12}(4), \frac{0}{12}(4), \frac{-6}{12}(2) $    \\[0.15em]  \hline &&&&&&&  \\[-1.15em]
 $\ominus=L$&  $  (++-)$ &$\frac{-4}{12}$&$ ~~\frac{+9}{12},~~~~~~\frac{+4}{12}$ &$\frac{21}{48}$&$\frac{-165}{48}$&$\frac{\pm 3}{12}[\frac{\pm 3}{12}] $ &$  \frac{+3}{12}(0), \frac{-3}{12}(0), \frac{-3}{12}(0), \frac{-9}{12}(0) $  \\ [0.1em]  \hline &&&&&&&  \\[-1.15em]
$\oplus=R$ &  $\color{red} (+++)$&  $\frac{-5}{12}$ &$ ~~\frac{+9}{12},~~~~~~\frac{+4}{12}$ &$\frac{21}{48}$&$\frac{-165}{48}$&
  $\frac{\pm 3}{12}[\frac{\pm 3}{12}] $&$  \frac{+2}{12}(2), \frac{-4}{12}(3), \frac{-4}{12}(3), \frac{-10}{12}(2)$  \\[0.1em]   \hline &&&&& && \\[-1.15em]
  $\oplus=R$ &   $ (+--)$& $0$ &$ ~~\frac{+9}{12},~~~~~~\frac{+4}{12}$ &$\frac{21}{48}$&$\frac{-165}{48}$ &
$\frac{\pm 3}{12}[\frac{\pm 3}{12}]  $ &$ \frac{+7}{12}(0), \frac{+1}{12}(0), \frac{+1}{12}(0), \frac{-5}{12}(0) $  \\[0.1em] \hline &&&&&&& \\[-1.15em]
$\oplus=R$ &   $\color{red} (-+-)$& $\frac{+1}{12}$ &$ ~~\frac{+9}{12},~~~~~~\frac{+4}{12}$ &$\frac{21}{48}$&$\frac{-165}{48}$ &
 $\frac{\pm 3}{12}[\frac{\pm 3}{12}]  $&$\frac{+8}{12}(3), \frac{+2}{12}(2), \frac{+2}{12}(2), \frac{-4}{12}(3) $   \\[0.1em] \hline &&&&&&&  \\[-1.25em]
$\oplus=R$&  $(--+)$  &  $\frac{+4}{12}$ &$ ~~\frac{+9}{12},~~~~~~\frac{+4}{12}$ &$\frac{21}{48}$&$\frac{-165}{48} $& $\frac{\pm 3}{12}[\frac{\pm 3}{12}]  $&  $  \frac{+11}{12}(0), \frac{+5}{12}(0), \frac{+5}{12}(0), \frac{-1}{12}(0)$ \\[0.15em] 
\hline
\end{tabular}
\end{center}
\caption{ One index spinor form for $V^3_{0}$: In the multiplicity, the order of $ \frac{\pm 3}{12}[\frac{\pm3}{12}]$ is  $\frac{+3}{12}[\frac{+3}{12}], \frac{-3}{12}[\frac{+3}{12}],  \frac{+3}{12}[\frac{-3}{12}],$ and $\frac{-3}{12}[\frac{-3}{12}]$.  Massless states are $ {\color{red} 4\left(\Psi_{[\alpha']L,1}+\Psi_{[\alpha']L,\bar1} \right)}  \oplus 20 \left( \Phi_{[\alpha' ] L, 1}+\Phi_{[\alpha' ] L, \bar1}\right) \oplus 20 \left( \Phi_{[\alpha' ] R, 1}+\Phi_{[\alpha' ] R, \bar1}\right)  $, where  multiplicity 2 of $T_3$ is taken into account.} \label{tab:T3ZeroOneInSpinor}
\end{table}

\item One index vector form  for $V^3 $:
 vector forms do not give massless states because $(P+3V_-)\cdot a_3\ne 0$. 
   
\end{itemize}

\subsubsection{Twisted sector $T_6$\,$(\delta ^6=0$)} 
    
\begin{itemize}
\item One  index spinor form  for $V^6 $: we have \footnote{Its CTP conjugate is provided by
  $P = (\underline{7+, 5+, 5+, 5+, 5+, 5+, 5+}; 5-)(-2,-2,-2,-2,0,-2,-4,-2)'$.}

\dis{
   &6V_0=\left(\frac{-5}{2},\,\frac{-5}{2},\,\frac{-5}{2},\,\frac{-5}{2},\,\frac{-5}{2},\,\frac{-5}{2},\,\frac{-5}{2} ;\, \frac{+5}{2}\right)
\left(\frac{4}{ 2},\, \frac{4}{ 2},\, \frac{4}{ 2},\,\frac{4}{ 2},\, 0,\, \frac{4}{ 2},\, \frac{7}{ 2},\, \frac{ 3}{ 2}\right)',~V^2_0=\frac{ 338}{144}\\[0.2em]
 &P  =(\underline{3+, 5+, 5+, 5+, 5+, 5+, 5+};5-)(-2,-2,-2,-2,0,-2,-3,-1)',~ \frac{-159}{12},\\[0.2em] 
&P +6V_0=\left(\underline{ -1, 0^6};0\right)\left(0^6,\frac{ 1}{2},\frac{ 1}{2} \right)',~(P+6V_0)\cdot a_3=0,
}
which saturate the needed masslessness condition $\frac32$ of $T_6$, which are tabulated in Table \ref{tab:MultiplicityT6}.   Note that to make the phase an integer times $\frac{1}{12}$, we choose $-p_{\rm vev}\cdot\phi_s$ as:  Since $-(6/2) (V_0^2-\phi_s^2)$ is even number times $1/24$,  we choose a vector $p_{\rm vev}$: $-p_{\rm vev}\cdot\phi_s=-[(-2)\times \frac{5}{12} +(-2)\times \frac{4}{12} +(0)\times \frac{1}{12} ]=\frac{18}{12}$. So, we used $\frac{18}{12}$ in the table.

 \begin{table}[!h]
\begin{center}
\begin{tabular}{|cc|c|c|ccc|c|c|   }
 \hline &&&&&&&  \\[-1.15em]
 Chirality  &   $\tilde s$& $-\tilde{s}\cdot\phi_s$& ${\rm -}p_{\rm vec}\cdot\phi_s ,~P\cdot V_0$& $(6/2)\phi_s^2$,&$  -(6/2) V_0^2 $,& $\Delta_{6}^N[\delta^6] $   &$\Theta_0\,( {\cal P}_6^N)$  \\[0.15em] \hline &&&&&&&   \\[-1.15em]
$\ominus=L$& $  \color{red} (---)$  &  $\frac{+5}{12}$ & $\frac{+18}{12},~~~~~\frac{-3}{12}$ &$\frac{21}{24} $  &$\frac{-169}{24}$ & $0[0]  $& $\frac{+6}{12}\,(2) $   \\ [0.1em] \hline &&&&&&& \\[-1.25em]
 $\ominus=L$ & $   (-++)$&  $0$ &$\frac{+18}{12},~~~~~\frac{-3}{12}$ &$\frac{21}{24}$&$ \frac{-169}{24}$ & 
  $0[0]  $ &$\frac{+1}{12}\,(0) $  \\[0.15em]  \hline  &&&&&&&  \\[-1.25em]
$\ominus=L$&  $ \color{red}(+-+)$  & $\frac{-1}{12}$ &$ \color{red}\frac{+18}{12},~~~~~\frac{-3}{12}$ &$\frac{21}{24}$&$\frac{-165}{24}$& 
 $0 [0] $ &$\frac{0}{12}\,(4) $  \\[0.15em]  \hline &&&&&&&  \\[-1.25em]
 $\ominus=L$&  $ (++-)$ &$\frac{-4}{12}$&$\frac{+18}{12},~~~~~\frac{-3}{12}$ &$\frac{21}{24}$&$\frac{-165}{24}$&
$0[0]   $ &$\frac{-3}{12}\,(0) $  \\ [0.1em]  \hline &&&&&&& \\[-1.25em]
$\oplus=R$ &  $ \color{red} (+++)$&  $\frac{-5}{12}$ &$\frac{+18}{12},~~~~~\frac{-3}{12}$ &$\frac{21}{24}$&$\frac{-165}{24}$ & $0 [0]  $&$\frac{-4}{12}\,(3) $   \\[0.1em]  \hline &&&&& && \\[-1.25em]
$\oplus=R$ &   $ (+--)$& $0$ &$\frac{+18}{12},~~~~~ \frac{-3}{12}$ &$\frac{21}{24}$&$\frac{-165}{24}$ &$0[0] $ & $ \frac{+1}{12}\,(0) $   \\[0.1em] \hline &&&&&&&  \\[-1.25em]
$\oplus=R$ & $\color{red}(-+-)$&  $\frac{+1}{12}$ &$\frac{+18}{12},~~~~~\frac{-3}{12}$ &$\frac{21}{24}$&$\frac{-165}{24}$ &$0[0]  $&$ \frac{+2}{12}\,(2)$   \\[0.1em] \hline &&&&&&  &  \\[-1.25em]
$\oplus=R$&  $ (--+)$  &  $\frac{+4}{12}$ &$\frac{+18}{12},~~~~~\frac{-3}{12}$ &$\frac{21}{24}$&$\frac{-165}{24}$& $0[0]  $&$\frac{+5}{12}\,(0)$  \\[0.15em] 
\hline
\end{tabular}
\end{center}
\caption{One index  spinor from $V_0^6$:   Chiralities and multiplicities, ${\color{red}    \Psi_{[A]L}} \oplus  5\left(\Phi_{[A]L} +\Phi_{[A]R}  \right)$. } \label{tab:MultiplicityT6}
\end{table}
\item One index vector form  for $V^6_{0}$: 
 vector type forms cannot satisfy the masslessness condition.
 
\end{itemize}
  
The remaining SU(7) non-singlet massless particles together with SU(4)$'$ nonsinglets are presented in Appendix. The SU(7) and SU(4)$'$ indices are represented by $A$ and $\alpha'$, respectively.  Therefore, twisted sectors $T_3,T_6,$ and $T_9$ may  be guessed that they are not affected by Wilson lines. However, $T_3$ and $T_9$ are affected by Wilson lines because in the calculation of the phase $\Delta_k^0$ there is an additional factor $\frac{k}{2}$ (viz. Eq. (\ref{eq:PhaseMore})). Indeed, the inclusion of this factor $\frac{k}{2}$ correctly produces a combination of an anomaly free set. 

Let us comment on the multiplicities in $T_3$ and $T_6$. In $T_3$, the multiplicity is 2 as mentioned in subsubsection $T_3$. In $T_6$, we note that it is a $\Z_2$ shift which is  in fact an untwisted sector. The multiplicity of $\Z_2$ untwisted sector is 2, but it must include antiparticles also. Thus, the multiplicity of  $\Z_2$ untwisted sector is 1 \cite{ChoiKS03}. It is taken into account in the twisted sector $\Z_6$.

 \begin{table}[!t]
\begin{center} 
\begin{tabular}{|c|cc|c|c|ccccccc|  }
   \hline &&&&&  &&&&&&  \\[-1.15em]
 &~${\cal P}\times$(rep.) & Sector& ~Weight~ & $V_a^k$ &$Q_1$& $Q_2$&$Q_3$&$Q_4$
 &$Q_5$&$Q_6$& $Q_7$   \\[0.15em]   
  \hline
 &&&&&  &&&&&&  \\[ -1.05em]
$(a)$ & $\Psi^{[ABC]}_R$&  $U_1 $  &  $\left(\underline{----+++};+\right)(0^8)'$ & $0 $  &  $\frac{-6}{12}$& $\frac{6}{12}$& 0&0&0&0&0    \\ [0.3em] \hline &&&&&  &&&&&&  \\ [-1.05em]
$(b)$ & $2\,\Psi^{[AB]}_R$ & $T_3 $  & $\left(\underline{\frac{ 3}4\, \frac{ 3}4\, \frac{ -1}4\, \frac{-1}4\, \frac{ -1}4\, \frac{ -1}4\, \frac{ -1}4}; \frac{ 1}4 \right)\left(0^6\, \frac{-1}{4}\, \frac{-1}4  \right)'$ &$V_0^3$ & $\frac{3}{12}$& $\frac{ 3}{12}$& 0&0 & 0& $\frac{-3 }{12}$& $\frac{ -3}{12}$   \\[0.3em]  \hline &&&&&  &&&&&&  \\[-1.05em]
$(c)$ &$~8\,\Psi_{[A] R } $ & $T_3$& $\left(\underline{\frac{ -3}4\, \frac{1}4\, \frac{ 1}4\, \frac{1}4\, \frac{ 1}4\, \frac{ 1}4\, \frac{ 1}4}; \frac{- 1}4 \right)\left(0^6\, \frac{-1}{4}\, \frac{-1}4  \right)'$ &$V_0^3$ & $\frac{9}{12} $ &$\frac{-3}{12} $ & 0& 0& 0&$\frac{-3}{12} $ & $\frac{-3}{12} $   \\[0.3em] \hline &&&&& &&&&&&  \\[-1.05em]
$(d)$ &$\Psi_{[A]R }$ & $T_5^+$& $\left(\frac{11}{12} \frac{-1}{12} \frac{-1}{ 12} \frac{ -1}{12} \frac{ -1}{12} \frac{-1}{12} \frac{-1}{12}; \frac{-3}{12}\right)
\left( 0 \,0 \, 0 \, 0\,  0  \frac{4}{12} \frac{-3}{12} \frac{-3}{12}\right)'$ &$V_+^5$  &$ \frac{5}{12}$ &  $\frac{-3}{12}$&0&0&$\frac{4}{12}$&$\frac{-3}{12}$ &  $\frac{-3}{12}$    \\[0.3em] \hline &&&&&  &&&&&&  \\[-1.05em]
$(e)$ &$\Psi^{[A]}_R$ & $T_6 $& $\left(\underline{ -1, 0^6};0\right)\left(0^6,\frac{ 1}{2},\frac{ 1}{2} \right)'$ &$V_0^6$ & $ \frac{-12}{12}$&0& 0&0 & 0&$\frac{6}{12} $& $\frac{6}{12} $    \\[0.1em] \hline &&&&&  &&&&&& \\[-1.05em]
$(f)$ &$40 \left(\Phi_{[A]R }+\Phi^{[A]} _{R } \right) $&  $T_3$  &$\left(\underline{\frac{- 3}4\, \frac{ 1}4\, \frac{ 1}4\, \frac{ 1}4\, \frac{ 1}4\, \frac{ 1}4\, \frac{ 1}4}; \frac{ -1}4 \right)\left(0^6\, \frac{-1}{4}\, \frac{-1}4  \right)' \oplus$~H.c. &$V_0^3$ & 0 & $0$&0&0 &$0$&$0$&0  \\[0.3em] 
\hline &&&&&   &&&&&&  \\[-1.05em]
$(g)$ &$5\left(\Phi_{[A]R}+\Phi^{[A]}_R\right)$&  $T_6$  & $\left(\underline{ -1\, 0^6}\,0\right)\left(0^6\,\frac{ 1}{2}\, \frac{1}{2} \right)'\oplus$~H.c. &$V_0^6$ & 0&0 &0&0&0 &0& 0   \\[0.3em] 
\hline &&&&&  &&&&&& \\[-1.05em]
$(h)$ &$ 10\left( \Phi_{[A]R }+\Phi^{[A]}_{R }\right) $&  $T_5^+$  & $\left(\frac{11}{12} \frac{-1}{12} \frac{-1}{ 12} \frac{ -1}{12} \frac{ -1}{12} \frac{-1}{12} \frac{-1}{12}; \frac{-3}{12}\right)
\left( 0 \,0 \, 0 \, 0\,  0  \frac{4}{12} \frac{-3}{12} \frac{-3}{12}\right)'\oplus$~H.c. &$V_+^5$ & 0& 0&0&0 &0&0& 0  \\[0.25em] 
\hline
 &&&& & &&&&&&  \\[-1.05em] 
  & & && $\sum_i$ & $\frac{35}{12}$ &$\frac{63}{12}$&0&0 &$\frac{4}{12}$&$\frac{-51}{12}$&$\frac{-51}{12}$  \\[0.3em] 
\hline\hline &&&& &  &&&&&&  \\[-1.05em]
$(a')$ & $\Psi_{[\alpha']R} $&  $U_3$  &  $(0^8)\left(\underline{-+++};-+++\right)'$ & $0 $  &$0$  & $0$&$\frac{ 1}{12}$&$\frac{ -1/2}{12}$&$\frac{1/2}{12}$&$\frac{1/2}{12}$&$\frac{ 1/2}{12}$   \\ [0.3em] 
\hline &&&&&  &&&&&  \\ [-1.05em]
$(b')$ & $ \Psi^{[\alpha']}_R $&  $T_1^0$  & $\left( (\frac{1}{12})^7\,  ;\frac{-1}{12} \right)\left(\underline{ \frac{10}{12}\, \frac{-2}{12}\,\frac{-2}{12}\,\frac{-2}{12}};\frac{-6}{12}\,\frac{-2}{12}\,\frac{1}{12};\frac{-3}{12} \right)' $  &$V_0^1 $ & $\frac{7}{12}$&$\frac{-1}{12}$ & $\frac{4}{12}$&$\frac{-6}{12}$&$\frac{-2}{12}$ &$\frac{1}{12}$&$\frac{-3}{12}$   \\[0.3em] 
\hline &&&&&  &&&&&& \\[-1.05em]
$(c')$ & $ \Psi_{[\alpha'] R }$&  $T_4^0$  & $\left( (\frac{-1}{6})^7\,  ;\frac{1}{6} \right)\left(\underline{ \frac{-1}{3}\, \frac{1}{3}\,\frac{1}{3}\,\frac{1}{3}};0\,\frac{ 1}{3}\,\frac{ 1}{3}\,0 \right)' $  &$V_0 ^4$ & $\frac{-14}{12}$&$\frac{2}{12}$& $\frac{8}{12}$& 0&$\frac{4}{12}$ &$\frac{4}{12}$& 0   \\[0.3em] 
\hline &&&&&   &&&&&& \\[-1.05em]
$(d')$ & $ \Psi^{[\alpha']}_R $&  $T_5^0$  & $\left(  (\frac{1}{12})^7\,;\frac{-1}{12} \right)\left(\underline{ \frac{10}{12}\, \frac{-2}{12}\,\frac{-2}{12}\,\frac{-2}{12}};\frac{-6}{12}\,\frac{-2}{12}\,\frac{-5}{12}\,\frac{ 3}{12} \right)' $  &$V_0^5 $ & $\frac{7}{12}$&$\frac{-1}{12}$ &$\frac{4}{12}$&$\frac{-6}{12}$&$\frac{-2}{12}$ &$\frac{-5}{12}$&$\frac{ 3}{12}$  \\[0.3em] 
\hline &&&&&  &&&&&& \\[-1.05em]
$(e')$ &$ 10\left(\Phi_{[\alpha']R}+\Phi^{[\alpha']}_R\right)$&  $T_1^0$  &  H.c.$\oplus\left(  (\frac{1}{12})^7\,;\frac{-1}{12} \right)\left(\underline{ \frac{10}{12}\, \frac{-2}{12}\,\frac{-2}{12}\,\frac{-2}{12}};\frac{-6}{12}\,\frac{-2}{12}\,\frac{1}{12};\frac{-3}{12} \right)' $ &$V_0^1 $   &  0& 0&0&0 &0&0& 0  \\[0.3em] 
\hline &&&&&   &&&&&& \\[-1.05em]
$(f')$ &$ 5\left(\Phi_{[\alpha']R}+\Phi^{[\alpha']}_R\right)$&  $T_4^0$  & $\left(( \frac{-1}{6})^7\,  ;\frac{1}{6} \right)\left(\underline{ \frac{-1}{3}\, \frac{1}{3}\,\frac{1}{3}\,\frac{1}{3}};0\,\frac{ 1}{3}\,\frac{ 1}{3}\,0 \right)'\oplus $~H.c.  &$V_0 ^4$ & 0&0&0 &0&0& 0 &0  \\[0.3em] 
\hline &&&&&  &&&&&& \\[-1.05em]
$(g')$ &$10\left(\Phi_{[\alpha']R}+\Phi^{[\alpha']}_R\right) $&  $T_5^0$  &  H.c.$\oplus\left( (\frac{1}{12})^7\,  ;\frac{-1}{12} \right)\left(\underline{ \frac{10}{12}\, \frac{-2}{12}\,\frac{-2}{12}\,\frac{-2}{12}};\frac{-6}{12}\,\frac{-2}{12}\,\frac{-5}{12}\,\frac{ 3}{12} \right)' $  &$V_0^5 $   & 0& 0&0&0 &0&0& 0  \\[0.3em] 
\hline &&&&& &&&&&& \\[-1.05em]
$(h')$ &$ 7\left(\Phi_{[\alpha']R}+\Phi^{[\alpha']}_R\right)$&  $T_6 $  & $\left( 0^8\right)\left(\underline{ 1\, 0\,0\,0};0\,0\,\frac{ -1}{2}\,\frac{ -1}{2} \right)'\oplus $~H.c.  &$V_0 ^4$   & 0& 0&0&0 &0&0& 0  \\[0.3em] 
\hline &&&&& &&&&&& \\[-1.05em]
  & & && $\sum_i$ & $0$ &$0$&$\frac{17}{12}$&$\frac{-25/2}{12}$ &$\frac{1/2}{12}$&$\frac{1/2}{12}$&$\frac{1/2}{12}$  \\[0.25em] 
\hline
\end{tabular}
\end{center}
\caption{Non-singlet \antiSD~spectra represented as R-handed chiral fields. H.c. means the opposite numbers of those in the same site.  } \label{tab:SU7Spectra}
\end{table}
The SU(7) and SU(4)$'$ non-singlet massless states are summarized as R-handed fields in Table \ref{tab:SU7Spectra}. The matter fields are denoted as $\Psi$ and vectorlike representations are represented by $\Phi$. Some of $\Phi$ fields develop VEVs. The  $\Phi$ fields can be removed at the GUT scale if correct combinations of sectors and oscillators are satisfied. The chiral fields of Table \ref{tab:SU7Spectra} are
\dis{
\Psi^{[ABC]}_R\oplus 2\,\Psi^{[AB]}_R\oplus 8\,\Psi_{[A]R}  
  }  
which do not have the SU(7) nonabelian anomaly. In the untwisted sector, there is no $[\,\bar1\,]$. Thus, the family $\Psi^{[ABC]}$ from $U$ has more suppressed $Q_{\rm em}=\frac23$ quark Yukawa coupling and $\Psi^{[ABC]}$ is interpreted to include the 1st family members.   All $\Psi_{[A]R}$'s appear in twisted sectors.  Two chiral fields, $(e)$ and one combination from $(c)$, form a vectorlike pair and removed at a high energy scale. The field in $(d)$ and the remaining 7 fields from $(c)$ are the needed 8 fields for $\Psi_{[A]R}$.  $\Psi_{[A]R}$ from $(e)$, \ie from $T_5^+$,  is interpreted as $u^c$ because it can lead to the smallest Yukawa coupling among $Q_{\rm em}=\frac23$ quarks. The other 7 fields  $\Psi_{[A]R}$ from $T_3$ have the same fate.  Then, note that $t^c$ and $c^c$ are located in $T_3$.

Since both $\Psi^{[AB]}$ and $t^c$ (from $\Psi_{[A]}$) arise at $T_3$, the cubic Yukawa coupling is possible from the BEH boson from $T_6$. There are many possibilities for assigning $H_u$ and $H_d$ of the MSSM in $T_3$ and $T_6$. We will choose a specific one in Subsec. \ref{subsec:Yukawa}.
 
\subsection{U(1) charges and anomalous U(1)}\label{subsec:Uones}

We use the normalization that the index $\ell$ for fundamental representation {\bf N} of SU($N$) is 1. Then, the indices of some representations are \cite{RamondGroup},
\dis{
 {\rm SU}(N): &~\ell({\bf N})=1,~\ell\left({ [2]}  \right)=N-2,~ \ell\left([3] \right) = \frac{(N-2)(N-3)}{2},\\ 
&~  \ell({\rm Adj.})=2N,~\ell\left(  \{ 2\}  \right) =N+2, \\ 
&~ \ell\left(  \{ 3\}  \right) = \frac{(N+2)(N+3)}{2},\\[0.4em]
{\rm U(1)_{em}}: &~\ell(Q_{\rm em})=2Q_{\rm em}^2 . \label{eq:indices}
 } 
 where $ [2]$ means the dimension $\left({\bf \frac{N(N-1)}{2!}} \right)$ with two antisymmetric indices,  $\{ 2 \}$ means the dimension $\left({\bf \frac{N(N+1)}{2!}} \right)$ with two symmetric indices, $ [3]$ means the dimension $\left({\bf \frac{N(N-1)(N-2)}{3!}} \right)$ with three antisymmetric indices,  etc. For SU(7), the index of $\Psi^{[ABC]}$ is 10 and the index of $\Psi^{[AB]}$ is 5. We need these numbers for the contribution of $\Psi^{[ABC]}$ and $\Psi^{[AB]}$ to the U(1)-SU(7)$^2$ anomalies.
 
We choose the following seven U(1) directions, in terms of $Q_i$ \cite{LNP696},
\dis{
Q_1=\left( 1\,1\,1\,1\,1\,1\,1\,0\right)\left(0^8 \right)' & \\[0.2em]
Q_2=\left(0\,0\,0\,0\,0\,0\,0\,1 \right)\left(0^8 \right)' &;~~ Q_2'= \frac{103}{35}Q_1+Q_6+Q_7,\\[0.2em]
Q_3= \left( 0^8\right)\left(1\,1\,1\,1\,0\,0\,0\,0 \right)'&;~~ Q_3'= \frac{103}{63}Q_2+Q_6+Q_7,\\[0.2em]
Q_4= \left( 0^8\right)\left(0\,0\,0\,0\,1\,0\,0\,0 \right)'&;~~ Q_4'=\frac{7}{34}Q_3+Q_5,  \\[0.2em]
Q_5=\left( 0^8\right)\left(0\,0\,0\,0\,0\,1\,0\,0 \right)' &;~~ Q_5'=-\frac{7}{25}Q_4+Q_5 , \\[0.2em]
Q_6=\left( 0^8\right)\left(0\,0\,0\,0\,0\,0\,1\,0 \right)' &;~~ Q_6'= \frac{11}{255}  Q_4+Q_5 +\frac{2}{51}(Q_6+Q_7),\\[0.2em]
Q_7=\left(0^8 \right)\left(0\,0\,0\,0\,0\,0\,0\,1 \right)'  &,  \label{eq:Qorig}
}
where the redefined primed U(1) combinations give the identical sum for the SU(7) and SU(4)$'$ anomalies. Note that $Q_6'$ itself is anomaly free. In terms of $Q_2', \cdots, Q_5'$, we can redefine anomaly free combinations. Six nonabelian-anomaly-free U(1) combinations are denoted with tilde, 
\dis{
&\tilde{Q}_1=Q_2'- Q_3' ,\\ 
&\tilde{Q}_2=2(Q_2' + Q_3')-Q_4'  ,\\
&\tilde{Q}_3= Q_4'-Q_5' ,\\
&\tilde{Q}_4=  2Q_2'+\frac12 Q_4' -Q_5' , \\ 
&\tilde{Q}_5=Q_6'=  \frac{11}{255}  Q_4+Q_5 +\frac{2}{51}(Q_6+Q_7),  \\[0.2em] 
 &\tilde{Q}_6=  Q_6-Q_7.
}
The remaining U(1) must carry anomaly, which can be represented as
\dis{
Q_a &= Q_2'+aQ_3'+bQ_4'+cQ_5' .\label{eq:Qanomal}
}
Parameters $a,b$ and $c$ are determined by how one breaks the \antiSD, which defines the electroweak hypercharges or the electromagnetic charges of the SM particles. Note that \GG\,subgroup of SO(10) use the U(1) direction, or equivalently the $B-L$ direction  $(1\,1\,1\,0\,0\,0\,0;0)(0^8)'$. The \flip\,subgroup of SO(10) use instead  the U(1) direction $(1\,1\,1\,1\,1\, 0\,0;0)(0^8)'$.  \antiSD\,uses $(1\,1\,1\,1\,1\, 0\,0;0)(0^8)'$ and in addition   $(1\,1\,1\,0\,0\,1\,1;0)(0^8)'$. These two directions of \antiSD~can be fixed only after \antiSD\,is broken down to the SM gauge group. The U(1)$_X$ of \antiSD~is given by
\dis{
Q_X=Q_1+Q_2-\frac13(Q_3+Q_4) +Q_5+Q_6+Q_7,
}
which is anomaly-free. The orthogonalities of $Q_a$ with $Q_X$ and the above two \antiSD\,directions determine three parameters of (\ref{eq:Qanomal}).

\subsection{Yukawa couplings}\label{subsec:Yukawa}

For Yukawa couplings, we must satisfy all the symmetries of low energy effective fields 
and the selection rules in the orbifold compactification. For the fields from twisted sectors, the Yukawa coupling structure is simpler than those involving the untwisted sector fields. Consider for example a vectorlike set from $T_6$ in Table \ref{tab:MultiplicityT6}. For the coupling,
$\Phi^{[A]}_R\cdot\Phi_{[A]R} $, we must satisfy the selection rules for the right-mover and for the left-mover conditions. For the right-mover condition, 36 times $p_{\rm vec}\cdot\phi_s$ is  0 mod. integer. It is satisfied for the coupling $\Phi^{[A]}_R\cdot\Phi_{[A]R} $. For the left-mover condition, 36 times $P\cdot V$ is  0 mod. integer. It is also satisfied for the coupling $\Phi^{[A]}_R\cdot\Phi_{[A]R} $.

As commented above,  $t,c,t^c$ and  $c^c$ quarks are located  $T_3$. On the other hand, $u$ is located in $U$ and $u^c$ is located in $T_5^+$. Order 1 Yukawa coupling of the form ${\bf 21}(T_3)\times \overline{\bf 7}(T_3)\times\overline{\bf 7}(T_6)$ is possible if $H_u$  in $T_6$ is not removed at the GUT scale. This requires a hierarchy of scales,
\dis{
M_s\ll M_3
}
where $M_3$ is a vacuum expectation value of a singlet {\bf 1} in $T_3$,
\dis{
 \langle{\one} (T_3)\rangle=M_3.\label{eq:hierarchyM}
}
Eight $\Psi_{[A]}$'s and five  $(\Phi^{[A]}+\Phi_{[A]})$'s have the following Yukawa couplings
\dis{
{\one} (T_3)\Psi_{[A]}^i \Phi^{[A]}_\mu,~M_s\Phi_{[A]}^i\Phi^{[A]}_\mu;~i=1,\cdots,8,~\mu=1,\cdots,5.
}
Due to the hierarchy (\ref{eq:hierarchyM}), five $\Phi^{[A]}$'s of $T_6$ are paired with five  $\Psi_{[A]}$'s from $T_3$. Three $\Psi_{[A]}$'s of $T_3$ and five $\Phi_{[A]}$'s of $T_6$ remain light at this stage. Introducing an angle $\tan\theta=\frac{M_s}{M_3}$, five 
BEH fields 
\dis{
\cos\theta\Phi_{[A]}^\mu-c_{\mu i}\,\sin\theta \Psi_{[A]}^i.\label{eq:mixedBEH}
}
obtain mass of order $M_3\sin\theta$.  Because of the democracy of couplings, four out of five 
$\Phi_{[A]}$'s of $T_6$ remain light. Collecting light $\overline{\bf 7}$'s up to this stage, we have
\dis{
{\rm Three~}\Psi_{[A]}(T_3),~ \Psi_{[A]}(T_5^+),~{\rm Four~}\Phi_{[A]}(T_6).
}
At the \antiSD~level, still we have eight light $\overline{\bf 7}$'s. Thus, the BEH fields are located at $T_6$. Depending on $\theta$, the BEH fields contain  small components from $T_3$, viz. Eq. (\ref{eq:mixedBEH}). We interpret this angle as the ratio $m_c/m_t=\tan\theta$.
The $t$-quark Yukawa coupling in \antiFD\,is
\dis{
T^{\bf 21}_{3}T^{\overline{\bf 7}}_{3}T^{\overline{\bf 7}}_{6,\rm BEH}~( t{ \rm~mass}).\label{eq:BEH3rdt}
}
The BEH fields giving mass to the $b$-quark are located in $T_3$. There are 40 $\Phi^{[A]}$ fields in $(f)$ of Table \ref{tab:SU7Spectra}. Because of the mass democracy, there can remain some light fields. Most of them will be removed when \antiSD~is broken, but we need one $\Phi^{[A]}$ for the $Q_{\rm em}=-\frac13$ quark masses.
For the $b$-quark mass, we need the coupling
 \dis{
\sim\frac{1}{M_s}\,T^{\bf 21}_{3}T^{\bf 21}_{3}T^{\bf 21}_{3,\rm BEH}T^{\bf 7}_{3,\rm BEH} .\label{eq:BEH3rdb}
}
Thus, the $b$-quark mass is expected to be much smaller than the $t$-quark mass, $ O( \langle T^{ \bf 21 }_{3,\rm BEH}\rangle \langle T^{ \bf 7 }_{3,\rm BEH} \rangle/ M_s \langle T^{ \bf 7 }_{6,\rm BEH}\rangle    )$, where $ \langle T^{ \bf 21 }_{3,\rm BEH}\rangle$ is the SU(5) splitting VEV $\langle\Phi^{[67]}\rangle$. Thus, we expect $m_b/m_t\sim\frac{\langle\Phi^{[67]}\rangle}{M_s\tan\beta}$. Even if $\tan\beta=O(1)$, we can fit $m_b/m_t$ to the observed value by appropriately tuning $\langle\Phi^{[67]}\rangle$. A similar suppression occurs for the second family members.

 For the 1st family members, the story is different. This is because $d^c$ appears in {\bf 35} of \antiSD, appearing in $U$. The $d$-quark mass may arise  from
 \dis{
\sim\frac{1}{M_s^2} {\bf 35}_{U_1} {\bf 35}_{U_1}{\bf 7}_{T_{3}, \rm BEH} \langle {\one}_{T_5,\rm BEH}\rangle \langle {\one}_{T_6,\rm BEH}\rangle.\label{eq:dmass}
 }
 Let us check whether this coupling is present. $p_{\rm spin}\cdot\phi_s=\frac{10}{12}$ for ${\bf 35}_{U_1} {\bf 35}_{U_1}$. From Table \ref{tab:T3ZeroOneInSpinor}, we note $p_{\rm vec}\cdot\phi_s=\frac{-9}{12}$ for ${\bf 7}_{T_,BEH}  $. We need the remaining singlet combinations to provide $\frac{-1}{12}$. Since we do not list singlets here, it cannot be shown at this stage, but there are numerous singlets and we assume that it is possible. For the left-movewr conditions which are the gauge invariance conditions, the above coupling satisfies the condition. For the $u$-quark mass, we must consider a higher dimensional operator than Eq. (\ref{eq:dmass}),
 \dis{
\sim\frac{1}{M_s^3} {\bf 35}_{U_1}  {\overline{\bf 7}} _{T_5^+} {\overline{\bf 7}}_{T_6,\rm BEH} \langle {\one}_{T_5^-,\rm BEH}\rangle  \langle {\one}_{T_5^0,\rm BEH}\rangle \langle {\one}_{T_2^0,\rm BEH}\rangle.
 }
 It is because $u^c$  appears in $T_5^+$, requiring another field carrying  another Wilson shift $-$ to remove the Wilsone shift +. 
Thus, there exists a possibility that $m_u<m_d$. It is a new mechanism for the inverted 1st family quark mass structure.

\subsection{Missing partner mechanism}
\label{subsec:Missingpartner}

At the \antiSD~level, we assume that one pair $\Phi_{[A]}=\overline{\bf 7}$ (giving mass to the $t$ quark)  and $\Phi^{[A]}= {\bf 7}$ (giving mass to the $b$ quark) are survibing down to low energy. The missing partner mechanism is discussed in this setup.
   
In a sense, the absence of
$ {\bf 7}_{\rm BEH}\cdot\overline{\bf 7}_{\rm BEH}$ is not guaranteed at field theory level. In the MSSM, it is related to the $\mu$ problem, ``Why there does not exists  $H_uH_d$ at the GUT scale'' \cite{KimNilles84}. Some interesting solutions with hidden-sector quarks exist \cite{CKN92,CM93}. These solutions are based on the Peccei-Quinn (PQ) symmetry with the very light axion \cite{Baer15}. In the effective SUSY framework language, the superpotential $W$ should not allow $\mu H_uH_d$ by assigning a nonvanishing PQ quantum number  to the combination $H_uH_d$.  But, the global PQ symmetry is spoiled by gravity \cite{Barr92}. We may resort to some discrete subgroup, e.g. matter parity \cite{Ibanez92}, of a U(1) gauge symmetry \cite{KraussWilczek}.  Suppose assigning the mother gauge charges of $H_u$ and $H_d$ as $Q(H_u)=Q(H_d)=1$ such that the matter parity forbids $H_uH_d$ at the GUT scale. But we must allow the $t$-quark mass at the cubic order. It means, $t$ and $t^c$ carry the mother gauge charge, $Q(t)=Q(t^c)=-\frac12$ for example. In string compactification, we do not worry the gravity spoil of global symmetries. Just string selection rules are enough to consider the coupling.  It has been noted that some string compactifications do not lead to quadratic term in $W$ as in $\Z_3$ \cite{CM93}, but in non-prime orbifolds the absence of   $\overline{\bf 7} \cdot {\bf 7} $ must be studied case by case.  In our example discussed above, the coupling $ {\bf 7}_{\rm BEH}\cdot\overline{\bf 7}_{\rm BEH}$ is not allowed because ${\bf 7}_{\rm BEH}$ is located in $T_3$, and $\overline{\bf 7}_{\rm BEH}$ is located in $T_6$.  But, GUTs need the doublet-triplet splitting  that in the same GUT scale BEH multiplet the colored fields are superheavy while $H_u$ and $H_d$ remain light. In the absence of the coupling $ {\bf 7}_{\rm BEH}\cdot\overline{\bf 7}_{\rm BEH}$, 
the missing partner mechanism of \antiSD\, is realized. Consider  the coupling,
\dis{
&\frac{1}{M_s}\, \epsilon^{ABCDEFG}\, \Phi_{[AB]}\Phi_{[CD]}\Phi_{[EF]} \Phi_{[G]},~{\rm and/or} \\ 
&\frac{1}{M_s^2}\, \epsilon^{ABCDEFG}\, \Phi_{[AB]}\Phi_{[CD]} \Phi_{[E]}  \langle \Phi'_{[F]} \rangle  \langle \Phi''_{[G]} \rangle  , \label{eq:mPartner}
}
where $\Phi' $ and $\Phi''$ obtain string scale VEVs, and $\Phi_{[AB]}=\Phi_{[45]}$ of Eq. (\ref{eq:mPartner}) are essential for separating the color and weak parts. The BEH bosons $H_u$ and $H_d$ are in $\Phi_{[A]}$ and $\Phi^{[A]}$, respectively. Equation (\ref{eq:mPartner}) makes colored scalars heavy, viz.
\dis{
&\Phi^{[23]} \Phi^{[1]} \,\langle\Phi^{[45]}\rangle \,\langle\Phi^{[67]}\rangle  , \\
&\Phi^{[23]} \Phi^{[1]} \,\langle\Phi^{[45]}\rangle \langle \Phi'^{[6]} \rangle  \langle \Phi''^{[7]} \rangle  , \label{eq:missingPartner}
}
where $\Phi^{[1]}$ is $Q_{\rm em}=-\frac13$ colored boson whose partner is $\Phi^{[23]}$. The color-weak separating VEV $\langle\Phi^{[45]}\rangle\approx \Mg\sim M_s$ is the key making the colored scalar heavy. In the same multiplet $\Phi^{[A]}$, the BEH doublet $H_d$ is present at $\Phi^{[4]}$ and $\Phi^{[5]}$. But, the indices 4 and 5 are already used for the GUT scale VEV, hence $H_d$ does not find a partner in $\Phi^{[AB]}$. This is the missing partner mechanism we realize in \antiSD.

\section{Conclusion}\label{sec:Conclusion}

In this paper, we proposed unification of families in string compactification. Here, we suggested anti-SU($N$) scheme \cite{Kim80} where the adjoint representation of SU($N$) is not needed for breaking the GUT group down to SU(3)$_c\times$U(1)$_{\rm em}$. It is pointed out that the anti-SU($N$) scheme has a merit in string compactification and can even save the SO(32) heterotic string theory for many phenomenological purposes. 
  The minimal model for UGUTF  is SU(7)  GUT with the representation $ [\,3\,]+ 2\,[\,2\,]+8\,[\,\bar{1}\,]$. We show it explicitly that this representation is realized in the $\Z_{12-I}$ orbifold compactification of $\EE8$ heterotic string.
  
The large top quark mass is possible in  \antiSD, where $t^c$ is in $\overline{\bf 7}=[\,\bar{1}\,]$. In the example discussed, $\overline{\bf 7}$'s in $T_3$ contain $t^c$ and $c^c$. The cubic coupling $\Psi^{[AB]}_{R,T_3} \Psi_{[A]R,T_3}\Phi_{[B]R,T_6,BEH}$ gives a dimension-3 superpotential for the $t$ mass. This cubic coupling is the only possible dimension 3 superpotential in our model and hence only $m_t$ is expected to be of order the electroweak scale. Other fermion masses are much smaller than $m_t$. We also presented an argument why there is an inverted mass ratio in the u-quark family. It is because $u^c$ is located in $T_5^+$ which requires another Wilson line shifted singlet field.
Finally, we presented the missing partner mechanism in string compactification. The key assumption, the absence of the coupling ${\bf 7}_{\rm BEH}\cdot\overline{\bf 7}_{\rm BEH}$ in the superpotential, is achieved here by locating ${\bf 7}_{\rm BEH}$ and $\overline{\bf 7}_{\rm BEH}$ separately  in $T_3$ and  $T_6$.   Then, it is shown that the missing partner mechanism works for $H_u$ and $H_d$ in $\overline{\bf 7}_{\rm BEH}$ and ${\bf 7}_{\rm BEH}$. The colored particles in $\overline{\bf 7}_{\rm BEH}$ and ${\bf 7}_{\rm BEH}$ find their partners in $\overline{\bf 21}_{\rm BEH}$ and ${\bf 21}_{\rm BEH}$ and obtain superheavy masses.

 Here, we neglected the details of singlet vacuum expectation values, toward removing vectorlike representations, though the  singlet VEVs  have been widely used  in other string compactification papers toward the MSSM \cite{KKK07}. In this sense,  the \antiSD~presented in this paper may be   an aethetic choice toward a desirable UGUTF. Other physics implications such as the quark and lepton mass textures, dark matter, and very light axions, including SU(7) singlet representations, will be presented elsewhere.

\acknowledgments{This work is supported in part by the National Research Foundation (NRF) grant funded by the Korean Government (MEST) (No. 2005-0093841) and by the IBS (IBS-R017-D1-2014-a00). }

\section*{Appendix: \antiSD~GUT in  $\Z_{12-I}$}\label{sec:Appendix}
In this Appendix, we list up the remaining non-singlet states not included in Sec. \ref{sec:SU7Model} where  $U, T_3$, and $T_6$ are discussed.  We only show the sectors containing SU(7) or SU(4)$'$ nonsinglets. They are listed up in the order of $T_4, T_1, T_2,$ and $T_{5}$.
     
\subsubsection{Twisted sector $T_1\, \left(  \delta_{\rm vec}^1=\frac{1}{12}\right)$}
 The masslessness condition for $ 2{c}_1$ requires $(p_{\rm vec}+ \phi_s)^2=\frac{66}{144}-$(oscillator contributions). Oscillator conribution from the right mover is $2\delta^1=\frac{24}{144}=2\cdot\frac{1}{12}$.

\begin{itemize} 
  
\item   One  index spinor form  for $V^1_{0}$:  
for the spinor,
\dis{
&V_0=\left(\frac{-5}{12},\,\frac{-5}{12},\,\frac{-5}{ 12},\,\frac{-5}{12},\,\frac{-5}{12},\,\frac{-5}{12},\,\frac{-5}{12} ;\, \frac{+5}{12}\right)
\left(\frac{4}{12},\, \frac{4}{12},\, \frac{4}{12},\,\frac{4}{12},\, 0,\, \frac{4}{12},\, \frac{7}{12},\, \frac{3}{12}\right)',~V^2_0=\frac{338}{144}\\[0.2em]
&P=\left(\underline{-++++++}; -\right)
\left(--------\right)',~P\cdot V_0=\frac{-30}{12},\\[0.2em]
&P+V_0=\left(\underline{\frac{-11}{12},\,\frac{1}{12},\,\frac{1}{ 12},\,\frac{ 1}{12},\,\frac{1}{12},\,\frac{1}{12},\,\frac{1}{12}} ;\, \frac{-1}{12}\right)
\left( \frac{-2}{12},\, \frac{-2}{12},\, \frac{-2}{12},\, \frac{-2}{12},\, \frac{-6}{12},\, \frac{-2}{12},\,  \frac{1}{12},\, \frac{-3}{12}\right)' , ~{ (P+V_0)\cdot a_3\ne 0 },\label{eq:T1eq0}
}
we have $(P+V_0)^2=\frac{194}{144}$.   The masslessness condition for $2\tilde{c}_1$ requires $(P+ V^a)^2=\frac{210}{144}-$(oscillator contributions). Note that  $(1/2)\phi_s^2-(1/2) V_0^2=-\frac{148}{12}=-13+\frac{+2/3}{12}$. The oscillator contribution is $\frac{16}{144}=2\cdot \frac{2/3}{12}$. So we need $-\Delta_4^N$ to cancel $\frac{+2/3}{12}$ in $(1/2)\phi_s^2-(1/2) V_0^2$. 
These are shown in Table
\ref{tab:T1OneZeroVecSpin}.
 
 \begin{table}[!h]
\begin{center}
\begin{tabular}{|cc|c|c|ccc|c|c| }
 \hline &&&&&&& \\[-1.15em]
  Chirality  &   $\tilde s$& $-\tilde{s}\cdot\phi_s$& $-p_{\rm vec}\cdot \phi_s,P\cdot V_0$& $(1/2)\phi_s^2$,&$  -(1/2) V_0^2 $,& $\Delta_{1}^N[\delta^1],
  \Delta_{1}^N[-\delta^1]  
  $   &$\Theta_0\,( {\cal P}_4^N)$    \\[0.15em]
 \hline &&&&&&&  \\[-1.15em]
$\ominus=L$& $(---)$  &  $\frac{+5}{12}$ & $~~\frac{0}{12},~~~~~~\frac{+6}{12}$ &$\frac{21}{72} $  &$\frac{-457}{72}$ & $ \frac{-2/3}{12}[\frac{+1}{12}],\frac{-2/3}{12}[\frac{-1}{12}]$& $\frac{+12}{12}\,(4),\frac{+10}{12}\,(2)$  \\ [0.1em] \hline &&&&&&& \\[-1.25em]
 $\ominus=L$ & $ \color{red}  (-++)$&  $0$ &$~~\frac{0}{12},~~~~~~\frac{+6}{12}$ &$\frac{21}{72} $  &$  \frac{-169}{72}$& $\frac{+2/3}{12}[\frac{+1}{12}],\frac{+2/3}{12}[\frac{-1}{12}] $& $ \frac{+7}{12}\,(0),\frac{+5}{12}\,(0)$   \\[0.15em]  \hline  &&&&&&&  \\[-1.25em]
$\ominus=L$&  $ (+-+)$  & $\frac{-1}{12}$ &$~~\frac{0}{12},~~~~~~\frac{+6}{12}$ &$\frac{21}{72} $  &$\frac{-457}{72}$& $\frac{-2/3}{12}[\frac{+1}{12}],\frac{-2/3}{12}[\frac{-1}{12}]$& $ \frac{+6}{12}\,(2),\frac{+4}{12}\,(3)$   \\[0.15em]  \hline &&&&&&&  \\[-1.25em]
 $\ominus=L$&  $\color{red} (++-)$ &$\frac{-4}{12}$&$~~\frac{0}{12},~~~~~~\frac{+6}{12}$ &$\frac{21}{72} $  &$\frac{-457}{72}$&$\frac{-2/3}{12}[\frac{+1}{12}],\frac{-2/3}{12}[\frac{-1}{12}]$& $ \frac{+3}{12}\,(0),\frac{+1}{12}\,(0)$   \\[0.1em]  \hline &&&&&&& \\[-1.25em]
$\oplus=L$& $(+++)$  &  $\frac{-5}{12}$ & $~~\frac{0}{12},~~~~~~\frac{+6}{12}$ &$\frac{21}{72} $  &$\frac{-457}{72}$ & $ \frac{-2/3}{12}[\frac{+1}{12}],\frac{-2/3}{12}[\frac{-1}{12}]$& $\frac{+2}{12}\,(2),\frac{0}{12}\,(4)$  \\ [0.1em] \hline &&&&&&& \\[-1.25em]
 $\oplus=L$ & $  \color{red} (+--)$&  $0$ &$~~\frac{0}{12},~~~~~~\frac{+6}{12}$ &$\frac{21}{72} $  &$ \frac{-457}{72}$& $\frac{-2/3}{12}[\frac{+1}{12}],\frac{-2/3}{12}[\frac{-1}{12}] $& $\frac{+7}{12}\,(0),\frac{+5}{12}\,(0)$   \\[0.15em]  \hline  &&&&&&&  \\[-1.25em]
$\oplus=L$&  $ (-+-)$  & $\frac{+1}{12}$ &$~~\frac{0}{12},~~~~~~\frac{+6}{12}$ &$\frac{21}{72} $  &$\frac{-457}{72}$& $\frac{-2/3}{12}[\frac{+1}{12}],\frac{-2/3}{12}[\frac{-1}{12}]$& $ \frac{+8}{12}\,(3),\frac{+6}{12}\,(2)$   \\[0.15em]  \hline &&&&&&&  \\[-1.25em]
 $\oplus=L$&  $\color{red} (--+)$ &$\frac{+4}{12}$&$~~\frac{0}{12},~~~~~~\frac{+6}{12}$ &$\frac{21}{72} $  &$\frac{-457}{72}$&$\frac{-2/3}{12}[\frac{+1}{12}],\frac{-2/3}{12}[\frac{-1}{12}]$& $\frac{+11}{12}\,(0),\frac{+9}{12}\,(0)$   \\[0.15em] 
\hline
\end{tabular}
\end{center}
\caption{One index spinor from $V_0^1$: Chiralities and multiplicities are $ 11\left( \Phi_{[A]   L} +\Phi_{[A]   R}\right)$. } \label{tab:T1OneZeroVecSpin}
\end{table}

\end{itemize} 
  
\begin{itemize} 
 
\item One index vector form  for $V^2_{0}$: we have 

\dis{
&2V_0=\left(\frac{-5}{6},\,\frac{-5}{6},\,\frac{-5}{6},\,\frac{-5}{6},\,\frac{-5}{6},\,\frac{-5}{6},\,\frac{-5}{6} ;\, \frac{+5}{6}\right)
\left(\frac{4}{6},\, \frac{4}{6},\, \frac{4}{6},\,\frac{4}{6},\, 0,\, \frac{4}{6},\, \frac{7}{6},\, \frac{ 3}{6}\right)',~V^2_0=\frac{ 338}{144},\\[0.2em]
&P =( {1,1,1,1,1,1,1};-1)(\underline{0,-1,-1,-1},0,-1, -1,  -1)' ,~P\cdot V_0=\frac{-65}{12}\\[0.2em]
&P +2V_0=\left(  {\frac{1}{6},\,\frac{1}{6},\,\frac{1}{6},\,\frac{1}{6},\,\frac{1}{6},\,\frac{1}{6},\,\frac{1}{6} } ; \frac{-1}{6}\right)
\left(  \underline{\frac{2}{3},\, \frac{-1}{3},\, \frac{-1}{3},\, \frac{-1}{3}},\,0,\,   \frac{-1}{3},\,\frac{1}{6},\, \frac{-1}{2}\right)',~{ (P_2+2V_0)\cdot a_3\ne 0.}
}
Note that $(P +2V_0)^2=\frac{50}{36}=\frac{216}{144}-\frac{16}{144}$, which needs $\frac{16}{144}=2\cdot\frac{2/3}{12}$  as an oscillator contribution with $\Delta_2^N=\frac{2/3}{12}$.  We have $(2/2)\phi_s^2-(2/2)V_0^2=-\frac{296}{144}=-2-\frac{2/3}{12}$. So, we need $+\Delta_2^N$. Massless states are presented in Table \ref{tab:T2ZeroVec}.

 \begin{table}[!h]
\begin{center}
\begin{tabular}{|cc|c|c|ccc|c|c| }
 \hline &&&&&&& \\[-1.15em]
 Chirality  &   $\tilde s$& $-\tilde{s}\cdot\phi_s$& ${\rm -}p_{\rm vec}\cdot\phi_s ,~P\cdot V_0$& $(2/2)\phi_s^2$,&$  -(2/2) V_0^2 $,& $\Delta_{2}^N[\delta^2],\Delta_{2}^N[-\delta^2] $   &$\Theta_0\,( {\cal P}_2^N)$    \\[0.15em]
 \hline &&&&&&&  \\[-1.15em]
$\ominus=L$& $   (---)$  &  $\frac{+5}{12}$ & $\frac{+9}{12},\frac{-3}{12}$ &$\frac{21}{24} $  &$\frac{-169}{24}$ & $\frac{+2/3}{12}[\frac{+2}{12}],\frac{+2/3}{12}[\frac{-2}{12}]  $& $\frac{+1}{12}\,(0),\frac{+9}{12}\,(0) $  \\ [0.1em] \hline &&&&&&& \\[-1.25em]
 $\ominus=L$ & $   (-++)$&  $0$ &$\frac{+9}{12},\frac{-3}{12}$ &$\frac{21}{24}$&$ \frac{-169}{24}$ & 
  $\frac{+2/3}{12}[\frac{+2}{12}],\frac{+2/3}{12}[\frac{-2}{12}]  $ &$\frac{+8}{12}\,(3),\frac{+4}{12}\,(3) $  \\[0.15em]  \hline  &&&&&&&  \\[-1.25em]
$\ominus=L$&  $ \color{red}(+-+)$  & $\frac{-1}{12}$ &$\frac{+9}{12},\frac{-3}{12}$ &$\frac{21}{24}$&$\frac{-165}{24}$& 
 $\frac{+2/3}{12}[\frac{+2}{12}],\frac{+2/3}{12}[\frac{-2}{12}] $ &$\frac{+7}{12}\,(0),\frac{+3}{12}\,(0) $  \\[0.15em]  \hline &&&&&&&  \\[-1.25em]
 $\ominus=L$&  $ (++-)$ &$\frac{-4}{12}$&$\frac{+9}{12},\frac{-3}{12}$ &$\frac{21}{24}$&$\frac{-165}{24}$&
$\frac{+2/3}{12}[\frac{+2}{12}],\frac{+2/3}{12}[\frac{-2}{12}] $ &$\frac{+4}{12}\,(3),\frac{0}{12}\,(4) $   \\ [0.1em]  \hline &&&&&&& \\[-1.25em]
$\oplus=R$ &  $ \color{red} (+++)$&  $\frac{-5}{12}$ &$\frac{+9}{12},\frac{-3}{12}$ &$\frac{21}{24}$&$\frac{-165}{24}$ & $\frac{+2/3}{12}[\frac{+2}{12}],\frac{+2/3}{12}[\frac{-2}{12}]  $&$\frac{+3}{12}\,(0),\frac{-1}{12}\,(0) $   \\[0.1em]  \hline &&&&& &&  \\[-1.25em]
$\oplus=R$ &   $ (+--)$& $0$ &$\frac{+9}{12}, \frac{-3}{12}$ &$\frac{21}{24}$&$\frac{-165}{24}$ &$\frac{+2/3}{12}[\frac{+2}{12}],\frac{+2/3}{12}[\frac{-2}{12}] $ & $ \frac{+8}{12}\,(3),\frac{+4}{12}\,(3) $  \\[0.1em] \hline &&&&&&& \\[-1.25em]
$\oplus=R$ & $(-+-)$&  $\frac{+1}{12}$ &$\frac{+9}{12},\frac{-3}{12}$ &$\frac{21}{24}$&$\frac{-165}{24}$ &$\frac{+2/3}{12}[\frac{+2}{12}],\frac{+2/3}{12}[\frac{-2}{12}] $&$\frac{+9}{12}\,(0),\frac{+5}{12}\,(0) $  \\[0.1em] \hline &&&&&&&  \\[-1.25em]
$\oplus=R$&  $ (--+)$  &  $\frac{+4}{12}$ &$\frac{+9}{12},\frac{-3}{12}$ &$\frac{21}{24}$&$\frac{-165}{24}$& $\frac{+2/3}{12}[\frac{+2}{12}],\frac{+2/3}{12}[\frac{-2}{12}] $&$\frac{+12}{12}\,(4),\frac{+8}{12}\,(3) $   \\[0.15em] 
\hline
\end{tabular}
\end{center}
\caption{One index  vector from $V_0^2$:   Chiralities and multiplicities, $   13\left(\Phi_{[A]L} +\Phi_{[A]R}  \right)$. } \label{tab:T2ZeroVec}
\end{table}
  
\end{itemize} 
 
\subsubsection{Twisted sector $T_5\, (\delta^5=\frac{1}{12})$}

\begin{itemize} 
   
\item  One index spinor form  for $V^5_{+}$:  
 
  \dis{
 &5V_+=\left( \frac{-5}{12},\, \frac{-5}{12},\, \frac{-5}{12},\, \frac{-5}{12},\, \frac{-5}{12},\, \frac{-5}{12},\, \frac{-5}{12} ;  \,\frac{45}{12}\right)
\left(0,\,0,\, 0,\, 0,\, 0,\, \frac{20}{12},\,  \frac{135}{12},\, \frac{-45}{12} \right)',~V_+^2 =\frac{914 }{144}, \\[0.2em]
&P=(\underline{-++++++ }; 7-)(0,0,0,0,0,-2,-11 , 4)',~P\cdot V_+=\frac{-375}{12}=\frac{-3}{12},  \\[0.2em]
&P+5V_+=\left(\underline{\frac{-11}{12},\,\frac{ 1}{12},\,\frac{1}{ 12},\,\frac{ 1}{12},\,\frac{ 1}{12},\,\frac{ 1}{12},\,\frac{ 1}{12}} ;\, \frac{ 3}{12}\right)
\left( 0,\,0,\, 0,\, 0,0, \frac{-4}{12},\,  \frac{3}{12},\, \frac{ 3}{12}\right)', ~  (P+5V_0)\cdot a_3\ne 0,\label{eq:T5qn}
}
which gives $(P+5V_+)^2=\frac{170}{144}$, and the oscillator contribution  of  $\frac{40}{144}=2\frac{5/3}{12} $  is needed.
Note that $(5/2)\phi_s^2  -(5/2) V_+^2=\frac{-18-4/3}{12}$, which means we select $ \Delta_{5}^N[\delta^5]$ and $ \Delta_{5}^N[ -\delta^5]$. The total gauge shift is even but it is even due to (odd shift) from  E$_8$ and (odd shift) from  E$_8'$. Thus, the gauge quantum numbers must be the opposite of Eq. (\ref{eq:T5qn}). These are  shown in Table
\ref{tab:T5PlusSpin}.

 \begin{table}[!h]
\begin{center}
\begin{tabular}{|cc|c|c|ccc|c|c| }
 \hline &&&&&&& \\[-1.15em]
  Chirality  &   $\tilde s$& $-\tilde{s}\cdot\phi_s$& $p_{\rm vec}\cdot\phi_s ,P\cdot V_+$& $(5/2)\phi_s^2$,&$  -(5/2) V_+^2 $,& $  \Delta_{4}^N[\delta^5],  \Delta_{5}^N[ -\delta^5]$   &$\Theta_0\,( {\cal P}_5^N)$    \\[0.15em]
 \hline &&&&&&&  \\[-1.05em]
$\ominus=L$& $   (---)$  &  $\frac{+5}{12}$ & $\frac{+18}{12},\frac{-3}{12}$ &$\frac{105}{144} $  &$\frac{-2285}{144}$ & $\frac{+5/3}{12}[\frac{+1}{12}],\frac{+5/3}{12}[\frac{-1}{12}]$& $\frac{+9}{12}\,(0),\frac{+7}{12}\,(0)   $  \\ [0.1em] \hline &&&&&&& \\[-1.05em]
 $\ominus=L$ & $  \color{red}(-++)$&  $0$ &$\frac{+18}{12},\frac{-3}{12}$ &$\frac{105}{144}$&$ \frac{-2285}{144}$ & 
  $\frac{+5/3}{12}[\frac{+1}{12}],\frac{+5/3}{12}[\frac{-1}{12}]$ &$\frac{+4}{12}\,(3),\frac{+2}{12}\,(2)    $  \\[0.15em]  \hline  &&&&&&&  \\[-1.25em]
$\ominus=L$&  $(+-+)$  & $\frac{-1}{12}$ &$\frac{+18}{12},\frac{-3}{12}$ &$\frac{105}{144} $  &$\frac{-2285}{144}$& 
 $\frac{+5/3}{12}[\frac{+1}{12}],\frac{+5/3}{12}[\frac{-1}{12}]$ &$\frac{+3}{12}\,(0),\frac{+1}{12}\,(0)   $  \\[0.15em]  \hline &&&&&&&  \\[-1.05em]
 $\ominus=L$&  $ \color{red}(++-)$ &$\frac{-4}{12}$&$\frac{+18}{12},\frac{-3}{12}$ &$\frac{105}{144} $  &$\frac{-2285}{144}$&
$\frac{+5/3}{12}[\frac{+1}{12}],\frac{+5/3}{12}[\frac{-1}{12}] $ &$\frac{0}{12}\,(4),\frac{-2}{12}\,(2)   $   \\ [0.1em]  \hline &&&&&&& \\[-1.05em]
$\oplus=R$ &  $ (+++)$&  $\frac{-5}{12}$ &$\frac{+18}{12},\frac{-3}{12}$ &$\frac{105}{144} $  &$\frac{-2285}{144}$& $\frac{+5/3}{12}[\frac{+1}{12}],\frac{+5/3}{12}[\frac{-1}{12}]$&$\frac{-1}{12}\,(0),\frac{-3}{12}\,(0)    $   \\[0.1em]  \hline &&&&& &&  \\[-1.05em]
$\oplus=R$ &   $ \color{red} (+--)$& $0$ &$\frac{+18}{12},\frac{-3}{12}$ &$\frac{105}{144} $  &$\frac{-2285}{144}$ &$\frac{+5/3}{12}[\frac{+1}{12}],\frac{+5/3}{12}[\frac{-1}{12}]$ &
 $\frac{+4}{12}\,(3),\frac{+2}{12}\,(2)    $  \\[0.1em] \hline &&&&&&& \\[-1.05em]
$\oplus=R$ & $(-+-)$&  $\frac{+1}{12}$ &$\frac{+18}{12},\frac{-3}{12}$ &$\frac{105}{144} $  &$\frac{-2285}{144}$ &$\frac{+5/3}{12}[\frac{+1}{12}],\frac{+5/3}{12}[\frac{-1}{12}]$&$\frac{+5}{12}\,(0),\frac{+3}{12}\,(0)    $  \\[0.1em] \hline &&&&&&&  \\[-1.05em]
$\oplus=R$&  $ \color{red} (--+)$  &  $\frac{+4}{12}$ &$\frac{+18}{12},\frac{-3}{12}$ &$\frac{105}{144} $  &$\frac{-2285}{144}$& $\frac{+5/3}{12}[\frac{+1}{12}],\frac{+5/3}{12}[\frac{-1}{12}] $&$\frac{+8}{12}\,(3),\frac{+6}{12}\,(2)   $   \\[0.15em] 
\hline
\end{tabular}
\end{center}
\caption{One index spinor from $V_+^5$: Chiralities and multiplicities, ${\color{red}\Psi^{[A]}_{L, 1} }\oplus  10\left(\Phi^{[A]}_{L, 1} +\Phi^{[A]}_{R, 1}  \right)$. } \label{tab:T5PlusSpin}
\end{table}

\end{itemize} 
 
\subsection{SU(4)$'$ spectra from twisted sectors $T$}

\subsubsection{Twisted sector $T_6$\,$(\delta^6=0)$}
   
\begin{itemize}

\item Hidden index vector form  for $V^6_{0}$: 
 we have  
\dis{
   &6V_0=\left(\frac{-5}{2},\,\frac{-5}{2},\,\frac{-5}{2},\,\frac{-5}{2},\,\frac{-5}{2},\,\frac{-5}{2},\,\frac{-5}{2} ;\, \frac{+5}{2}\right)
\left(\frac{4}{ 2},\, \frac{4}{ 2},\, \frac{4}{ 2},\,\frac{4}{ 2},\, 0,\, \frac{4}{ 2},\, \frac{7}{ 2},\, \frac{ 3}{ 2}\right)',~V^2_0=\frac{ 338}{144}\\[0.2em]
 &P  =(5+, 5+, 5+, 5+, 5+, 5+, 5+ ;5-)(\underline{-1,-2,-2,-2},0,-2,-3,-1)',~ P\cdot V_0=\frac{-160 }{12},\\[0.2em] 
&P +6V_0=\left( 0^8\right)\left(\underline{1,0,0,0},0,0,\frac{ 1}{2},\frac{ 1}{2} \right)',~(P+6V_0)\cdot a_3=0,
}
which saturates the needed masslessness condition $\frac32$ of $T_6$, which are tabulated in Table \ref{tab:T6OneHid}.   
$P$ is odd under both $E_8$ and $E_8'$, we complex conjugate the $E_8'$ quantum numbers.
 \begin{table}[!h]
\begin{center}
\begin{tabular}{|cc|c|c|ccc|c|c| }
 \hline &&&&&&& \\[-1.15em]
  Chirality  &   $\tilde s$& $-\tilde{s}\cdot\phi_s$& $-p_{\rm vec}\cdot\phi_s,P\cdot V_0$& $(6/2)\phi_s^2$,&$  -(6/2) V_0^2 $,& $\Delta_{6}^N[\delta^6]$   &$\Theta_0\,( {\cal P}_6^N)$    \\[0.15em]
 \hline &&&&&&&  \\[-1.15em]
$\ominus=L$& $ (---)$  &  $\frac{+5}{12}$ & $\frac{+18}{12},~~~~~\frac{-4}{12}$ &$\frac{21}{24} $  &$\frac{-169}{24}$ & $0 [0] $& $\frac{+5}{12}\,(0)$  \\ [0.1em] \hline &&&&&&& \\[-1.25em]
 $\ominus=L$ & $\color{red}(-++)$&  $0$ &$\frac{+18}{12},~~~~~\frac{-4}{12}$ &$\frac{21}{24}$&$  \frac{-169}{24}$ & 
  $0 [0] $ &$\frac{0}{12}\,(4) $  \\[0.15em]  \hline  &&&&&&&  \\[-1.25em]
$\ominus=L$&  $(+-+)$  & $\frac{-1}{12}$ &$\frac{+18}{12},~~~~~\frac{-4}{12}$ &$\frac{21}{24}$&$\frac{-165}{24}$& 
 $0 [0]$ &$\frac{-1}{12}\,(0)$ \\[0.15em]  \hline &&&&&&&  \\[-1.25em]
 $\ominus=L$&  $ \color{red} (++-)$ &$\frac{-4}{12}$&$\frac{+18}{12},~~~~~\frac{-4}{12}$ &$\frac{21}{24}$&$\frac{-165}{24}$&
$0 [0]$ &$\frac{-4}{12}\,(3) $   \\ [0.1em]  \hline &&&&&&& \\[-1.25em]
$\oplus=R$ &  $ (+++)$&  $\frac{-5}{12}$ &$\frac{+18}{12},~~~~~\frac{-4}{12}$ &$\frac{21} 
  {24}$&$\frac{-165}{24}$& $0 [0]$&$\frac{-5}{12}\,(0) $   \\[0.1em]  \hline &&&&& &&  \\[-1.25em]
$\oplus=R$ &   $ \color{red} (+--)$& $0$ &$\frac{+18}{12},~~~~~\frac{-4}{12}$ &$\frac{21}{24}$&$\frac{-165}{24}$ &$0 [0]$ &
 $\frac{0}{12}\,(4) $  \\[0.1em] \hline &&&&&&& \\[-1.25em]
$\oplus=R$ & $(-+-)$&  $\frac{+1}{12}$ &$\frac{+18}{12},~~~~~\frac{-4}{12}$ &$\frac{21}{24}$&$\frac{-165}{24}$ &$0 [0]$&$ \frac{+1}{12}\,(0)$  \\[0.1em] \hline &&&&&&&  \\[-1.25em]
$\oplus=R$&  $ \color{red} (--+)$  &  $\frac{+4}{12}$ &$\frac{+18}{12},~~~~~\frac{-4}{12}$ &$\frac{21}{24}$&$\frac{-165}{24}$& $0 [0]$&$\frac{+4}{12}\,(3)$   \\[0.15em] 
\hline
\end{tabular}
\end{center}
\caption{ Hidden  index spinor from $V_0^6$: Chiralities and multiplicities, $  7\left(\Phi^{[\alpha'] }_L +\Phi^{[\alpha'] } _R \right)$. } \label{tab:T6OneHid}
\end{table}
  
\end{itemize}
  
\subsubsection{Twisted sector $T_4 \, ( \delta^4=0 )$}
 
\begin{itemize} 

\item Hidden  index vector form  for $V^4_{0}$:   
\dis{
 & 4V_0 = \left(\frac{-5}{3},\,\frac{-5}{3},\,\frac{-5}{3},\,\frac{-5}{3},\,\frac{-5}{3},\,\frac{-5}{3},\,\frac{-5}{3} ;\, \frac{5}{3}\right)
\left( \frac{4}{3},\,\frac{4}{3},\,\frac{4}{3},\, \frac{4}{3},\,0,\,\frac{4}{3},\, \frac{7}{3},\,\frac{3}{3} \right)',~~V^2_0=\frac{338}{144},\\[0.2em]
 &P =\left( \frac{3}{2},\frac{3}{2},\frac{3}{2},\frac{3}{2},\frac{3}{2},\frac{3}{2},\frac{3}{2} ; \frac{-3}{2} \right)\left( \underline{-2,-1,-1,-1},0,-1,-2,-1\right)',~P_2\cdot V_0=\frac{-101  }{12}=\frac{-5}{12},\\[0.2em]
 &P +4V_0= \left( \frac{-1}{6},\frac{-1}{6},\frac{-1}{6},\frac{-1}{6},\frac{-1}{6},\frac{-1}{6},\frac{-1}{6} ; \frac{1}{6} \right)
\left(\underline{\frac{ -2}{3}, \frac{ 1}{3}, \frac{ 1}{3},  \frac{ 1}{3}},0, \frac{ 1}{3}, \frac{  1}{3}, 0 \right)',~(P_2+4V_0)\cdot a_3 \ne 0. 
\label{eq:T4Zero}
}
Since the shift is  $(\rm odd)\times(odd)$ under $\EE8$, we interchange the SU(4)$'$ gauge quantum numbers. Note  $(P_2+4V_0)^2=\frac{44}{36}=\frac{48}{36}-\frac{4 }{36}$. The oscillator contribution of $2\cdot\frac{2/3}{12}$ is needed. Using $-(4/2)(V_0^2-\phi_s^2)=\frac{-49-1/3}{12}$, we choose $-\Delta_4^N$. 

Table \ref{tab:T4ZeroVecHid}.

 \begin{table}[!h]
\begin{center}
\begin{tabular}{|cc|c|c|ccc|c|c| }
 \hline &&&&&&& \\[-1.15em]
  Chirality  &   $\tilde s$& $-\tilde{s}\cdot\phi_s$& $-p_{\rm vec}\cdot\phi_s,P\cdot V_0$& $(4/2)\phi_s^2$,&$  -(4/2) V_0^2 $,& $-\Delta_{4}^N[\delta^4]$   &$\Theta_0\,( {\cal P}_6^N)$    \\[0.15em]
 \hline &&&&&&&  \\[-1.15em]
$\ominus=L$& $\color{red} (---)$  &  $\frac{+5}{12}$ & $\frac{+18}{12},~~~~~\frac{-4}{12}$ &$\frac{21}{24} $  &$\frac{-169}{24}$ & $\frac{-2/3}{12}[0]$& $\frac{+12}{12}\,(4)$  \\ [0.1em] \hline &&&&&&& \\[-1.25em]
 $\ominus=L$ & $(-++)$&  $0$ &$\frac{+14}{12},~~~~~\frac{-5}{12}$ &$\frac{21}{36}$&$ \frac{-169}{36}$ & 
  $\frac{-2/3}{12}[0] $ &$\frac{+7}{12}\,(0) $  \\[0.15em]  \hline  &&&&&&&  \\[-1.25em]
$\ominus=L$&  $\color{red}(+-+)$  & $\frac{-1}{12}$ &$\frac{+18}{12},~~~~~\frac{-4}{12}$ &$\frac{21}{24}$&$\frac{-165}{24}$& 
 $\frac{-2/3}{12}[0]$ &$\frac{+6}{12}\,(2)$  \\[0.15em]  \hline &&&&&&&  \\[-1.25em]
 $\ominus=L$&  $ (++-)$ &$\frac{-4}{12}$&$\frac{+18}{12},~~~~~\frac{-4}{12}$ &$\frac{21}{24}$&$\frac{-165}{24}$&
$\frac{-2/3}{12}[0]$ &$\frac{+3}{12}\,(0) $   \\ [0.1em]  \hline &&&&&&& \\[-1.25em]
$\oplus=R$ &  $ \color{red}(+++)$&  $\frac{-5}{12}$ &$\frac{+18}{12},~~~~~\frac{-4}{12}$ &$\frac{21} 
  {24}$&$\frac{-165}{24}$& $\frac{-2/3}{12}[0]$&$\frac{+2}{12}\,(2) $   \\[0.1em]  \hline &&&&& &&  \\[-1.25em]
$\oplus=R$ &   $  (+--)$& $0$ &$\frac{+18}{12},~~~~~\frac{-4}{12}$ &$\frac{21}{24}$&$\frac{-165}{24}$ &$\frac{-2/3}{12}[0]$ &
 $\frac{+7}{12}\,(0) $  \\[0.1em] \hline &&&&&&& \\[-1.25em]
$\oplus=R$ & $\color{red}(-+-)$&  $\frac{+1}{12}$ &$\frac{+18}{12},~~~~~\frac{-4}{12}$ &$\frac{21}{24}$&$\frac{-165}{24}$ &$\frac{-2/3}{12}[0]$&$ \frac{+8}{12}\,(3)$  \\[0.1em] \hline &&&&&&&  \\[-1.25em]
$\oplus=R$&  $  (--+)$  &  $\frac{+4}{12}$ &$\frac{+18}{12},~~~~~\frac{-4}{12}$ &$\frac{21}{24}$&$\frac{-165}{24}$& $\frac{-2/3}{12}[0]$&$\frac{+11}{12}\,(0)$   \\[0.15em] 
\hline
\end{tabular}
\end{center}
\caption{Hidden sector vector from $V_0^4$: Chiralities and multiplicities, $ {\color{red}\Psi^{[\alpha']}_{L}}\oplus 5\left(\Phi^{[\alpha']}_{L}+\Phi^{[\alpha']}_{R} \right)$. } \label{tab:T4ZeroVecHid}
\end{table}
 
\end{itemize} 
   
\subsubsection{Twisted sector $T_1\, (\delta^1=\frac{1}{12})$} 
 
\begin{itemize} 
  
\item Hidden  index spinor form  for $V^1_{0}$:  
for the spinor,
\dis{
&V_0=\left(\frac{-5}{12},\,\frac{-5}{12},\,\frac{-5}{ 12},\,\frac{-5}{12},\,\frac{-5}{12},\,\frac{-5}{12},\,\frac{-5}{12} ;\, \frac{+5}{12}\right)
\left(\frac{4}{12},\, \frac{4}{12},\, \frac{4}{12},\,\frac{4}{12},\, 0,\, \frac{4}{12},\, \frac{7}{12},\, \frac{3}{12}\right)',~V^2_0=\frac{338}{144}\\[0.2em]
&P=\left( {+++++++}; -\right)
\left(\underline{+---},----\right)',~P\cdot V_0=\frac{-31}{12}=\frac{+5}{12},\\[0.2em]
&P+V_0=\left( {\frac{1}{12},\,\frac{1}{12},\,\frac{1}{ 12},\,\frac{ 1}{12},\,\frac{1}{12},\,\frac{1}{12},\,\frac{1}{12}} ;\, \frac{-1}{12}\right)
\left( \underline{\frac{10}{12},\, \frac{-2}{12},\, \frac{-2}{12},\, \frac{-2}{12}},\, \frac{-6}{12},\, \frac{-2}{12},\,  \frac{1}{12},\, \frac{-3}{12}\right)' , ~ (P+V_0)\cdot a_3\ne 0 ,\label{eq:T1eq0}
}
we have $(P+V_0)^2=\frac{170}{144}=\frac{210}{144}-\frac{40}{144}$. Reverse the gauge quantum numbers.  Note that  $(1/2)\phi_s^2-(1/2) V_0^2=-\frac{148}{12}=-12+\frac{-1/3}{12}$. The oscillator contribution is $\frac{40}{144}=2\cdot \frac{5/3}{12}$. So we need $-\Delta_4^N=\frac{-5/3}{12}$ to cancel $\frac{+2/3}{12}$ in $(1/2)\phi_s^2-(1/2) V_0^2$. 
These are shown in Table
\ref{tab:T1HidZeroSpin}.
 
 \begin{table}[!h]
\begin{center}
\begin{tabular}{|cc|c|c|ccc|c|c| }
 \hline &&&&&&& \\[-1.15em]
  Chirality  &   $\tilde s$& $-\tilde{s}\cdot\phi_s$& $-p_{\rm vec}\cdot \phi_s,P\cdot V_0$& $(1/2)\phi_s^2$,&$  -(1/2) V_0^2 $,& $\Delta_{1}^N[\delta^1],
  \Delta_{1}^N[-\delta^1]  
  $   &$\Theta_0\,( {\cal P}_4^N)$    \\[0.15em]
 \hline &&&&&&&  \\[-1.15em]
$\ominus=L$& $(---)$  &  $\frac{+5}{12}$ & $~~\frac{0}{12},~~~~~~\frac{+5}{12}$ &$\frac{21}{72} $  &$\frac{-169}{72}$ & $ \frac{-2/3}{12}[\frac{+1}{12}],\frac{-2/3}{12}[\frac{-1}{12}]$& $\frac{+9}{12}\,(0),\frac{+7}{12}\,(0)$  \\ [0.1em] \hline &&&&&&& \\[-1.25em]
 $\ominus=L$ & $ \color{red}  (-++)$&  $0$ &$~~\frac{0}{12},~~~~~~\frac{+5}{12}$ &$\frac{21}{72} $  &$ \frac{-169}{72}$& $\frac{-5/3}{12}[\frac{+1}{12}],\frac{-5/3}{12}[\frac{-1}{12}] $& $ \frac{+4}{12}\,(3),\frac{+2}{12}\,(2)$   \\[0.15em]  \hline  &&&&&&&  \\[-1.25em]
$\ominus=L$&  $ (+-+)$  & $\frac{-1}{12}$ &$~~\frac{0}{12},~~~~~~\frac{+5}{12}$ &$\frac{21}{72} $  &$\frac{-169}{72}$& $\frac{-2/3}{12}[\frac{+1}{12}],\frac{-2/3}{12}[\frac{-1}{12}]$& $\frac{+3}{12}\,(0),\frac{+1}{12}\,(0)$   \\[0.15em]  \hline &&&&&&&  \\[-1.25em]
 $\ominus=L$&  $\color{red} (++-)$ &$\frac{-4}{12}$&$~~\frac{0}{12},~~~~~~\frac{+5}{12}$ &$\frac{21}{72} $  &$\frac{-169}{72}$&$\frac{-2/3}{12}[\frac{+1}{12}],\frac{-2/3}{12}[\frac{-1}{12}]$& $\frac{0}{12}\,(4),\frac{-2}{12}\,(2)$   \\[0.1em]  \hline &&&&&&& \\[-1.25em]
$\oplus=L$& $(+++)$  &  $\frac{-5}{12}$ & $~~\frac{0}{12},~~~~~~\frac{+5}{12}$ &$\frac{21}{72} $  &$\frac{-169}{72}$ & $ \frac{-2/3}{12}[\frac{+1}{12}],\frac{-2/3}{12}[\frac{-1}{12}]$& $\frac{-1}{12}\,(0),\frac{-3}{12}\,(0)$  \\ [0.1em] \hline &&&&&&& \\[-1.25em]
 $\oplus=L$ & $  \color{red} (+--)$&  $0$ &$~~\frac{0}{12},~~~~~~\frac{+5}{12}$ &$\frac{21}{72} $  &$ \frac{-169}{72}$& $\frac{-2/3}{12}[\frac{+1}{12}],\frac{-2/3}{12}[\frac{-1}{12}] $& $\frac{+4}{12}\,(3),\frac{+2}{12}\,(2)$   \\[0.15em]  \hline  &&&&&&&  \\[-1.25em]
$\oplus=L$&  $ (-+-)$  & $\frac{+1}{12}$ &$~~\frac{0}{12},~~~~~~\frac{+5}{12}$ &$\frac{21}{72} $  &$\frac{-169}{72}$& $\frac{-2/3}{12}[\frac{+1}{12}],\frac{-2/3}{12}[\frac{-1}{12}]$& $ \frac{+5}{12}\,(0),\frac{+3}{12}\,(0)$   \\[0.15em]  \hline &&&&&&&  \\[-1.25em]
 $\oplus=L$&  $\color{red} (--+)$ &$\frac{+4}{12}$&$~~\frac{0}{12},~~~~~~\frac{+5}{12}$ &$\frac{21}{72} $  &$\frac{-169}{72}$&$\frac{-2/3}{12}[\frac{+1}{12}],\frac{-2/3}{12}[\frac{-1}{12}]$& $\frac{+8}{12}\,(3),\frac{+6}{12}\,(2)$   \\[0.15em] 
\hline
\end{tabular}
\end{center}
\caption{ Hidden index spinor from $V_0^1$: Chiralities and multiplicities are ${\color{red}\Psi_{[\alpha']L} } \oplus10\left( \Phi_{[\alpha']   L} +\Phi_{[\alpha']   R}\right)$. } \label{tab:T1HidZeroSpin}
\end{table}

\end{itemize} 
 
\subsubsection{Twisted sector $T_5\, (\delta^5=\frac{1}{12})$}

\begin{itemize} 
 
\item Hidden index vector form  for $V^5_{0}$: we have 

\dis{
&5V_0=\left(\frac{-25}{12},\,\frac{-25}{12},\,\frac{-25}{ 12},\,\frac{-25}{12},\,\frac{-25}{12},\,\frac{-25}{12},\,\frac{-25}{12} ;\, \frac{+25}{12}\right)
\left(\frac{20}{12},\, \frac{20}{12},\, \frac{20}{12},\,\frac{20}{12},\, 0,\, \frac{20}{12},\, \frac{35}{12},\, \frac{15}{12}\right)',~V^2_0=\frac{ 338}{144}\\[0.2em]
&P_2=( {2,2,2,2,2,2,2 }; -2)(\underline{5-,3-,3-,3-},+ ,3-,5-,3- )',~P\cdot V_0=\frac{-456}{12}=\frac{0}{12}, \\[0.2em]
&P_2+5V_0=\left(\underline{\frac{-1}{12},\,\frac{-1}{12},\,\frac{-1}{ 12},\,\frac{-1}{12},\,\frac{-1}{12},\,\frac{-1}{12},\,\frac{-1}{12}} ;\, \frac{+1}{12}\right)
\left( \underline{ \frac{-10}{12},\,  \frac{2}{12},\,\frac{2}{12} ,\, \frac{2}{12}},\,\frac{ 6}{12},\, \frac{2}{12},\,  \frac{5}{12},\, \frac{-3}{12}\right)',~ 
}
 which gives $(P_1+5V_0)^2=\frac{194}{144}=\frac{210}{144}-\frac{16}{144}$. The oscillator contribution of $2\cdot\frac{2/3}{12}$ is needed.     Massless states are shown in Table
Table \ref{tab:T5HidZeroSpin}.
 
 \begin{table}[!h]
\begin{center}
\begin{tabular}{|cc|c|c|ccc|c|c| }
 \hline &&&&&&& \\[-1.15em]
  Chirality  &   $\tilde s$& $-\tilde{s}\cdot\phi_s$& $p_{\rm vec}\cdot\phi_s ,P\cdot V_0$& $(5/2)\phi_s^2$,&$  -(5/2) V_0^2 $,& $  \Delta_{4}^N[\delta^5],  \Delta_{5}^N[ -\delta^5]$   &$\Theta_0\,( {\cal P}_5^N)$    \\[0.15em]
 \hline &&&&&&&  \\[-1.05em]
$\ominus=L$& $   (---)$  &  $\frac{+5}{12}$ & $\frac{+18}{12},~~~~\frac{0}{12}$ &$\frac{105}{144} $  &$\frac{- 845}{144}$ & $\frac{+5/3}{12}[\frac{+1}{12}],\frac{+5/3}{12}[\frac{-1}{12}]$& $\frac{+11}{12}\,(0),\frac{+9}{12}\,(0)   $  \\ [0.1em] \hline &&&&&&& \\[-1.05em]
 $\ominus=L$ & $  \color{red}(-++)$&  $0$ &$\frac{+18}{12},~~~~\frac{0}{12}$ &$\frac{105}{144}$&$  \frac{- 845}{144}$ & 
  $\frac{+2/3}{12}[\frac{+1}{12}],\frac{+2/3}{12}[\frac{-1}{12}]$ &$\frac{+6}{12}\,(2),\frac{+4}{12}\,(3)    $  \\[0.15em]  \hline  &&&&&&&  \\[-1.25em]
$\ominus=L$&  $(+-+)$  & $\frac{-1}{12}$ &$\frac{+18}{12},~~~~\frac{0}{12}$ &$\frac{105}{144} $  &$\frac{- 845}{144}$& 
 $\frac{+5/3}{12}[\frac{+1}{12}],\frac{+5/3}{12}[\frac{-1}{12}]$ &$\frac{+5}{12}\,(0),\frac{+3}{12}\,(0)   $  \\[0.15em]  \hline &&&&&&&  \\[-1.05em]
 $\ominus=L$&  $ \color{red}(++-)$ &$\frac{-4}{12}$&$\frac{+18}{12},~~~~\frac{0}{12}$ &$\frac{105}{144} $  &$\frac{- 845}{144}$&
$\frac{+5/3}{12}[\frac{+1}{12}],\frac{+5/3}{12}[\frac{-1}{12}] $ &$\frac{2}{12}\,(2),\frac{0}{12}\,(4)   $   \\ [0.1em]  \hline &&&&&&& \\[-1.05em]
$\oplus=R$ &  $ (+++)$&  $\frac{-5}{12}$ &$\frac{+18}{12},~~~~\frac{0}{12}$ &$\frac{105}{144} $  &$\frac{- 845}{144}$& $\frac{+5/3}{12}[\frac{+1}{12}],\frac{+5/3}{12}[\frac{-1}{12}]$&$\frac{+1}{12}\,(0),\frac{-1}{12}\,(0)    $   \\[0.1em]  \hline &&&&& &&  \\[-1.05em]
$\oplus=R$ &   $ \color{red} (+--)$& $0$ &$\frac{+18}{12},~~~~\frac{0}{12}$ &$\frac{105}{144} $  &$\frac{- 845}{144}$ &$\frac{+5/3}{12}[\frac{+1}{12}],\frac{+5/3}{12}[\frac{-1}{12}]$ &
 $\frac{+6}{12}\,(2),\frac{+4}{12}\,(3)    $  \\[0.1em] \hline &&&&&&& \\[-1.05em]
$\oplus=R$ & $(-+-)$&  $\frac{+1}{12}$ &$\frac{+18}{12},~~~~\frac{0}{12}$ &$\frac{105}{144} $  &$\frac{- 845}{144}$ &$\frac{+5/3}{12}[\frac{+1}{12}],\frac{+5/3}{12}[\frac{-1}{12}]$&$\frac{+7}{12}\,(0),\frac{+5}{12}\,(0)    $  \\[0.1em] \hline &&&&&&&  \\[-1.05em]
$\oplus=R$&  $ \color{red} (--+)$  &  $\frac{+4}{12}$ &$\frac{+18}{12},~~~~\frac{0}{12}$ &$\frac{105}{144} $  &$\frac{- 845}{144}$& $\frac{+5/3}{12}[\frac{+1}{12}],\frac{+5/3}{12}[\frac{-1}{12}] $&$\frac{+10}{12}\,(2),\frac{+8}{12}\,(3)   $   \\[0.15em] 
\hline
\end{tabular}
\end{center}
\caption{One index spinor from $V_0^5$: Chiralities and multiplicities, ${\color{red}\Psi_{[\alpha']L } }\oplus  10\left(\Phi_{[\alpha']L } +\Phi_{[\alpha']R }  \right)$. } \label{tab:T5HidZeroSpin}
\end{table}
 
\item Hidden index spinor form  for $V^5_{+}$:  
  \dis{
 &5V_+=\left( \frac{-5}{12},\, \frac{-5}{12},\, \frac{-5}{12},\, \frac{-5}{12},\, \frac{-5}{12},\, \frac{-5}{12},\, \frac{-5}{12} ;  \,\frac{45}{12}\right)
\left(0,\,0,\, 0,\, 0,\, 0,\, \frac{20}{12},\,  \frac{135}{12},\, \frac{-45}{12} \right)',~V_+^2 =\frac{914 }{144}, \\[0.2em]
&P=(+++++++ ; 7-)(\underline{1,0,0,0},0,-2,-11 , 4)',~P\cdot V_+=\frac{-376}{12}=\frac{- 4}{12},  \\[0.2em]
&P+5V_+=\left( {\frac{ 1}{12},\,\frac{ 1}{12},\,\frac{1}{ 12},\,\frac{ 1}{12},\,\frac{ 1}{12},\,\frac{ 1}{12},\,\frac{ 1}{12}} ;\, \frac{ 3}{12}\right)
\left( \underline{1,0,0,0},0, \frac{-4}{12},\,  \frac{3}{12},\, \frac{ 3}{12}\right)', ~  (P+5V_0)\cdot a_3\ne 0 \label{eq:T5qn}
}
which gives $(P+5V_+)^2=\frac{194}{144}=\frac{210}{144}-\frac{16}{144}$. The oscillator contribution of $2\cdot\frac{2/3}{12}$ is needed.    
Note that $(5/2)\phi_s^2  -(5/2) V_+^2=\frac{-181-2/3}{12}$, which means we select $ \Delta_{5}^N[\delta^5]$ and $ \Delta_{5}^N[ -\delta^5]$.  Massless states are  shown in Table
\ref{tab:T5HidPlusSpin}.

 \begin{table}[!h]
\begin{center}
\begin{tabular}{|cc|c|c|ccc|c|c| }
 \hline &&&&&&& \\[-1.15em]
  Chirality  &   $\tilde s$& $-\tilde{s}\cdot\phi_s$& $p_{\rm vec}\cdot\phi_s ,P\cdot V_+$& $(5/2)\phi_s^2$,&$  -(5/2) V_+^2 $,& $  \Delta_{4}^N[\delta^5],  \Delta_{5}^N[ -\delta^5]$   &$\Theta_0\,( {\cal P}_5^N)$    \\[0.15em]
 \hline &&&&&&&  \\[-1.05em]
$\ominus=L$& $   (---)$  &  $\frac{+5}{12}$ & $\frac{+18}{12},\frac{-4}{12}$ &$\frac{105}{144} $  &$\frac{-2285}{144}$ & $\frac{+2/3}{12}[\frac{+1}{12}],\frac{+2/3}{12}[\frac{-1}{12}]$& $\frac{+7}{12}\,(0),\frac{+5}{12}\,(0)   $  \\ [0.1em] \hline &&&&&&& \\[-1.05em]
 $\ominus=L$ & $  \color{red}(-++)$&  $0$ &$\frac{+18}{12},\frac{-4}{12}$ &$\frac{105}{144}$&$  \frac{-2285}{144}$ & 
  $\frac{+2/3}{12}[\frac{+1}{12}],\frac{+2/3}{12}[\frac{-1}{12}]$ &$\frac{+2}{12}\,(2),\frac{-2}{12}\,(2)    $  \\[0.15em]  \hline  &&&&&&&  \\[-1.25em]
$\ominus=L$&  $(+-+)$  & $\frac{-1}{12}$ &$\frac{+18}{12},\frac{-4}{12}$ &$\frac{105}{144} $  &$\frac{-2285}{144}$& 
 $\frac{+2/3}{12}[\frac{+1}{12}],\frac{+2/3}{12}[\frac{-1}{12}]$ &$\frac{+1}{12}\,(0),\frac{-3}{12}\,(0)   $  \\[0.15em]  \hline &&&&&&&  \\[-1.05em]
 $\ominus=L$&  $ \color{red}(++-)$ &$\frac{-4}{12}$&$\frac{+18}{12},\frac{-4}{12}$ &$\frac{105}{144} $  &$\frac{-2285}{144}$&
$\frac{+2/3}{12}[\frac{+1}{12}],\frac{+2/3}{12}[\frac{-1}{12}]$ &$\frac{-2}{12}\,(2),\frac{-6}{12}\,(2)   $   \\ [0.1em]  \hline &&&&&&& \\[-1.05em]
$\oplus=R$ &  $ (+++)$&  $\frac{-5}{12}$ &$\frac{+18}{12},\frac{-4}{12}$ &$\frac{105}{144} $  &$\frac{-2285}{144}$& $\frac{+2/3}{12}[\frac{+1}{12}],\frac{+2/3}{12}[\frac{-1}{12}]$&$\frac{-3}{12}\,(0),\frac{-7}{12}\,(0)    $   \\[0.1em]  \hline &&&&& &&  \\[-1.05em]
$\oplus=R$ &   $ \color{red} (+--)$& $0$ &$\frac{+18}{12},\frac{-4}{12}$ &$\frac{105}{144} $  &$\frac{-2285}{144}$ &$\frac{+2/3}{12}[\frac{+1}{12}],\frac{+2/3}{12}[\frac{-1}{12}]$ &
 $\frac{+2}{12}\,(2),\frac{-2}{12}\,(2)    $  \\[0.1em] \hline &&&&&&& \\[-1.05em]
$\oplus=R$ & $(-+-)$&  $\frac{+1}{12}$ &$\frac{+18}{12},\frac{-4}{12}$ &$\frac{105}{144} $  &$\frac{-2285}{144}$ &$\frac{+2/3}{12}[\frac{+1}{12}],\frac{+2/3}{12}[\frac{-1}{12}]$&$\frac{+3}{12}\,(0),\frac{-1}{12}\,(0)    $  \\[0.1em] \hline &&&&&&&  \\[-1.05em]
$\oplus=R$&  $ \color{red} (--+)$  &  $\frac{+4}{12}$ &$\frac{+18}{12},\frac{-4}{12}$ &$\frac{105}{144} $  &$\frac{-2285}{144}$& $\frac{+2/3}{12}[\frac{+1}{12}],\frac{+2/3}{12}[\frac{-1}{12}] $&$\frac{+6}{12}\,(2),\frac{+2}{12}\,(2)   $   \\[0.15em] 
\hline
\end{tabular}
\end{center}
\caption{Hidden sector  $V_+^5$: massless states are  $  8\left(\Phi^{[\alpha']}_{L } +\Phi^{[\alpha']}_{ R }  \right)$. } \label{tab:T5HidPlusSpin}
\end{table}
   
\end{itemize} 


\begin{thebibliography}{99}
\def\prp#1#2#3{{Phys.\,Rep.}  {\bf #1} (#3) #2}
\def\rmp#1#2#3{{ Rev. Mod. Phys.}  {\bf #1} (#3) #2}
\def\npb#1#2#3{{ Nucl.\,Phys.\,B}   {\bf #1} (#3) #2}
\def\plb#1#2#3{{Phys.\,Lett.\,B}   {\bf #1} (#3) #2}
\def\prd#1#2#3{{Phys.\,Rev.\,D}  {\bf #1} (#3) #2}
\def\prl#1#2#3{{ Phys.\,Rev.\,Lett.}   {\bf #1} (#3) #2}
\def\err#1#2#3{ {\bf #1}   {\bf #1} (#3) #2\,(E)}
\def\jhep#1#2#3{{ JHEP}   {\bf #1} (#3) #2}
\def\jcap#1#2#3{{ JCAP}   {\bf #1} (#3) #2}
\def\zp#1#2#3{{ Z.\,Phys.}  {\bf #1} (#3) #2}
\def\epjc#1#2#3{{ Euro.\,Phys.\,J.\,C}  {\bf #1} (#3) #2}
\def\jpg#1#2#3{{J.\,Phys.\,G}   {\bf #1} (#3) #2}
\def\ijmpd#1#2#3{{ Int.\,J.\,Mod.\,Phys.\,D}   {\bf #1} (#3) #2}
\def\mpla#1#2#3{{Mod.\,Phys.\,Lett.\,A}   {\bf #1} (#3) #2}
\def\apj#1#2#3{{Astrophys.\,J.}   {\bf #1} (#3) #2}
\def\nat#1#2#3{{Nature}   {\bf #1} (#3) #2}
\def\sjnp#1#2#3{{ Sov.\,J.\,Nucl.\,Phys.}  {\bf #1} (#3) #2}
\def\apj#1#2#3{{Astrophys.\,J.}   {\bf #1} (#3) #2}
\def\mnra#1#2#3{{ Mon.\,Not.\,Roy.\,Astron.\,Soc.}   {\bf #1} (#3) #2}
\def\jetpl#1#2#3{{JETP\,Lett.}   {\bf #1} (#3) #2}
\def\pthp#1#2#3{{Prog.\,Theor.\,Phys.}  {\bf #1} (#3) #2}
\def\jkps#1#2#3{{J.\,Korean\,Phys.\,Soc.}  {\bf #1} (#3) #2}
\def\dum#1#2#3{  {\bf #1} (#3) #2}

\def\ibid#1#2#3{{\it ibid.} {\bf #1} (#3) #2}
\def\err#1#2#3{\ {\bf #1} (#3) #2\,(E)}
\def\err#1#2#3{\ {\bf #1} (#3) #2\,(E)}
  
 \bibitem{GG74} H.\,M. Georgi and S.\,L. Glashow, \emph{Unity of all elementary particle forces}, \prl{32}{438}{1974} [doi:  10.1103/PhysRevLett.32.438].
  
 \bibitem{GQW74} H. Georgi, H.\,R. Quinn, and S. Weinberg, \emph{Hierarchy of interactions in unified gauge theories}, \prl{33}{451}{1974} [doi:  10.1103/PhysRevLett.33.451].
  
 \bibitem{PS73} J.\,C. Pati and Abdus Salam, \emph{Unified lepton-hadron symmetry and a gauge theory of the basic interactions}, \prd{8}{1240}{1973} [doi:  10.1103/PhysRevD.8.1240].
 
\bibitem{Georgi79} H. Georgi, \emph{Towards a grand unified theory of flavor}, \npb{156}{126}{1979} [doi:  10.1016/0550-3213(79)90497-8].

\bibitem{SO10} H. Georgi, \emph{The state of the art -- Gauge theories}, AIP Conf. Proc. 23 (1975) 575-582  [doi: 10.1063/1.2947450]; H. Fritzsch and P. Minkowski,  \emph{Unified interactions of leptons and hadrons}, Annals Phys. 93 (1975) 193  [doi: 10.1016/0003-4916(75)90211-0].

\bibitem{Kim80} J.\,E. Kim, \emph{A model of flavor unity}, \prl{45}{1916}{1980} [doi:  10.1103/PhysRevLett.45.1916].

\bibitem{Kim81} J.\,E. Kim, \emph{Flavor unity in SU(7): Low mass magnetic monopole, doubly charged lepton, and Q=5/3,−4/3 quarks}, \prd{23}{2706}{1981} [doi:   10.1103/PhysRevD.23.2706].

\bibitem{Dimopoulos81} S. Dimopoulos and F. Wilczek, \emph{The unity of the fundamental interactions}, Conference: C81-07-31 ed. A. Zichichi (Plenum, New York, 1983), p.817 [Proceedings, 19th Course of the International School of Subnuclear Physics, Erice, Italy, July 31 - August 11, 1981] [doi:  10.1007/978-1-4613-3655-6].

\bibitem{Missing} K.\,S. Babu and S.\,M. Barr, \emph{Natural gauge hierarchy in SO(10)}, \prd{50}{3529}{1994} [doi:  hep-ph/9402291].

\bibitem{Frampton79} P. Frampton, \emph{SU(N) grand unification with several quark - lepton generations}, \plb{88}{299}{1979} [doi: 10.1016/0370-2693(79)90472-6].

\bibitem{FramptonPRL79} P. Frampton and S. Nandi, \emph{SU(9) grand unification of flavor with three generations}, \prl{43}{1460}{1979} [doi:  10.1103/PhysRevLett.43.1460].

\bibitem{Gross85} D.\,J. Gross, J.\,A. Harvey, E.\,J. Martinec, and R. Rohm, \emph{The heterotic string}, \prl{54}{502}{1985} [doi: 10.1103/PhysRevLett.54.502].
  
\bibitem{Candelas85} 
P. Candelas, G.\,T. Horowitz, A. Strominger, and E. Witten, \emph{Vacuum configurations for superstrings}, \npb{258}{46}{1985} [doi: 10.1016/0550-3213(85)90602-9].
  
\bibitem{Dixon85} L.\,J. Dixon, J.\,A. Harvey, C. Vafa, and E. Witten, \emph{Strings on orbifolds}, \npb{261}{678}{1985} [doi: 10.1016/0370-2693(87)90066-9].
  
\bibitem{Ibanez87} L.\,E. Iba\~nez, H.\,P. Nilles, and F. Quevedo, \emph{Orbifolds and Wilson lines}, \plb{187}{25}{1987} [doi: 10.1016/0370-2693(87)90066-9].
   
\bibitem{IKNQ87} L.\,E. Iba\~nez, J.\,E. Kim, H.\,P. Nilles, and F. Quevedo, \emph{Orbifold compactifications with three families of SU(3)$\times$SU(2)$\times$U(1)$^n$}, \plb{191}{282}{1987} [doi: 10.1016/0370-2693(87)90255-3]; J.\,A. Casas and C. Munoz, \emph{Three generation SU(3)$\times$SU(2)$\times$U(1)$_Y$ models from orbifolds}, \plb{214}{63}{1988} [doi: 10.1016/0370-2693(88)90452-2].

\bibitem{DKN84} J.\,P. Derendinger, J.\,E. Kim, and D.\,V. Nanopoulos, \emph{Anti-SU(5)}, \plb{139}{170}{1984} [doi:10.1016/0370-2693(84)91238-3 ].

\bibitem{Barr82} S.\,M. Barr, \emph{A new symmetry breaking pattern for SO(10) and proton decay}, \plb{112}{219}{1982} [doi: 10.1016/0370-2693(82)90966-2].

\bibitem{Ellis89} I.  Antoniadis, J.\,R. Ellis, J\,.S. Hagelin, and D.\,V. Nanopoulos, \emph{The flipped SU(5)$\times$U(1) string model revamped}, \plb{231}{65}{1989} [doi: 10.1016/0370-2693(89)90115-9].
 
 \bibitem{KimKyae07} J.\,E. Kim and B. Kyae, \emph{Flipped SU(5) from $\Z_{12-I}$ orbifold with Wilson line}, \npb{770}{47}{2007} [arXiv: hep-th/0608086].
  
\bibitem{KimPRL13} J.\,E. Kim,  \emph{Natural Higgs-flavor-democracy solution of the $\mu$ problem of supersymmetry and the QCD axion}, \prl{111}{031801}{2013} [arXiv:1303.1822 [hep-ph]].

\bibitem{KimPLB13}  J.\,E. Kim, \emph{Abelian discrete symmetries $\Z_N$ and $\Z_{nR}$ from string orbifolds }, \plb{726}{450}{2013} [arXiv:1308.0344 [hep-th]].

\bibitem{FramptonPLB79} P. Frampton and S. Nandi,
\emph{Estimate of flavor number from SU(5) grand unification }, \plb{85}{225}{1979} [doi:  10.1016/0370-2693(79)90584-7].

  
\bibitem{LNP696}  K.-S. Choi and J.\,E. Kim, {\it Quarks and Leptons from Orbifolded Superstring},  Lecture Notes in Physics Vol. 696 (Springer-Verlag, 2006), Chapter 10 and Appendix D.  

 \bibitem{ChoiKS03} K.-S. Choi and J.\,E. Kim, \emph{$\Z_2$ orbifold
compactication of heterotic string and 6D SO(16) and E$_7\times$SU(2) 
flavor unification models}, \plb{552}{81}{2003} :  [arXiv:hep-th/0206099]. 
  
\bibitem{RamondGroup}  P. Ramond, {\it Group Theory} (Cambridge University Press, Cambridge, England, 2010), Appendix 2;   
K.-S. Choi, \emph{Extended gauge symmetries in F-theory}, \jhep{1002}{004}{2010} [arXiv:0910.2571 [hep-th]].

\bibitem{KimNilles84} J.\,E. Kim and H.\,P. Nilles, \emph{The $\mu$ problem and the strong CP problem}, \plb{138}{150}{1984} [doi: 10.1016/0370-2693(84)91890-2].

\bibitem{CKN92} E.\,J. Chun, J.\,E. Kim, and H.\,P. Nilles, \emph{A natural solution of the $\mu$ problem with a composite axion in the hidden sector}, \npb{370}{105}{1992} [doi:  10.1016/0550-3213(92)90346-D].

\bibitem{CM93} J.\,A. Casas and C. Mu\~noz, \emph{Natural solution to the $\mu$ problem}, \plb{306}{288}{1993} [arXiv: hep-ph/9302227].

\bibitem{Baer15} H. Baer, K.-Y. Choi, J.\,E. Kim, and L. Roszkowski, \emph{Dark matter production in the early Universe: beyond the thermal WIMP paradigm}, \prp{555}{1}{2015} [arXiv:1407.0017 [hep-ph]].

\bibitem{Barr92}
S.\,M. Barr and D. Seckel, \emph{Planck scale corrections to axion models},
     \prd{46}{539}{1992} [doi: 10.1103/PhysRevD.46.539]; 
M. Kamionkowski and J. March-Russell, \emph{Planck scale physics and the Peccei-Quinn mechanism},
          \plb{282}{137}{1992} [hep-th/9202003];
R. Holman, S.\,D.\,H. Hsu, T.\,W. Kephart, E. W. Kolb, R. Watkins, and L.\,M. Widrow, \emph{Solutions to the strong CP problem in a world with gravity },
     \plb{282}{132}{1992} [hep-ph/9203206];
B. A. Dobrescu, \emph{The Strong CP problem versus Planck scale physics 
},  \prd{55}{5826}{1997} [hep-ph/9609221].

\bibitem{Ibanez92} L.\,E. Ibanez and G.\,G. Ross, \emph{Discrete gauge symmetries and the origin of baryon and lepton number conservation in supersymmetric versions of the standard model}, \npb{368}{3}{1992} [doi: 10.1016/0550-3213(92)90195-H].

\bibitem{KraussWilczek}  L.\,M. Krauss and F. Wilczek, \emph{Discrete gauge symmetry in continuum theories}, \prl{62}{1221}{1989} [doi: 10.1103/PhysRevLett.62.1221].


\bibitem{Raby05} T. Kobayashi, S. Raby, and R.\,J. Zhang, \emph{Searching for realistic 4d string models with a Pati-Salam symmetry: Orbifold grand unified theories from heterotic string compactification on a $\Z_6$ orbifold}, \npb{704}{3}{2005} [arXiv: hep-ph/0409098 ].

\bibitem{KKK07}J.\,E. Kim, J.-H. Kim,  and B. Kyae, \emph{Superstring standard model from $\Z_{12-I}$ orbifold compactification with and without exotics, and effective R-parity }, \jhep{0706}{034}{2007} [arXiv: hep-ph/0702228]; J.\,H. Huh, J.\,E. Kim, and B. Kyae,  {SU(5)$_{\rm flip}$ x SU(5)$'$ \emph{from} $\Z_{12-I}$}, \prd{80}{115012}{2009} [arXiv: 0904.1108 [hep-ph]];  
  O. Lebedev, H.\,P. Nilles, S. Raby, S. Ramos-Sanchez, M. Ratz,
P.\,K.\,S. Vaudrevange, and A. Wingerter,
\emph{The heterotic road to the MSSM with R parity}, \prd{77}{046013}{2008} 
  [arXiv:0708.2691 [hep-th]]; 
O. Lebedev, H.\,P. Nilles,  S. Ramos-Sanchez, M. Ratz, and
P.\,K.\,S. Vaudrevange, \emph{Heterotic mini-landscape  (II): Completing the search for MSSM vacua in a $\Z_6$ orbifold}, \plb{668}{331}{2008} [arXiv:0807.4384 [hep-th]].
  
 
\end{thebibliography}
\end{document}